\documentclass{article}

\usepackage{arxiv}

\usepackage[utf8]{inputenc} 
\usepackage[T1]{fontenc}    
\usepackage{hyperref}       
\usepackage{url}            
\usepackage{booktabs}       
\usepackage{amsfonts}       
\usepackage{nicefrac}       
\usepackage{microtype}      
\usepackage{lipsum}
\usepackage{natbib}
\usepackage{graphicx}
\usepackage{longtable}
\usepackage{amsmath}

\title{Constraining subglacial processes from surface velocity observations using surrogate-based Bayesian inference}
\author{Douglas Brinkerhoff \\
       Department of Computer Science \\
       University of Montana \\
       Missoula, MT 59812 \\
       \texttt{douglas1.brinkerhoff@umontana.edu}
       \And
       Andy Aschwanden \\
       Geophysical Institute \\
       University of Alaska Fairbanks \\
       Fairbanks, AK 99775 \\
       \texttt{aaschwanden@alaska.edu}
       \And 
       Mark Fahnestock \\
       Geophysical Institute \\
       University of Alaska Fairbanks \\
       Fairbanks, AK 99775 \\
       \texttt{mfahnestock@alaska.edu}}

\begin{document}
\maketitle

\begin{abstract}
Basal motion is the primary mechanism for ice flux outside Antarctica, yet a widely applicable model for predicting it in the absence of retrospective observations remains elusive.  This is due to the difficulty in both observing small-scale bed properties and predicting a time-varying water pressure on which basal motion putatively depends.  We take a Bayesian approach to these problems by coupling models of ice dynamics and subglacial hydrology and conditioning on observations of surface velocity in southwestern Greenland to infer the posterior probability distributions for eight spatially and temporally constant parameters governing the behavior of both the sliding law and hydrologic model.  Because the model is computationally expensive, classical MCMC sampling is intractable.  We skirt this issue by training a neural network as a surrogate that approximates the model at a sliver of the computational cost.  We find that surface velocity observations establish strong constraints on model parameters relative to a prior distribution and also elucidate correlations, while the model explains 60\,\% of observed variance.  However, we also find that several distinct configurations of the hydrologic system and stress regime are consistent with observations, underscoring the need for continued data collection and model development. 
\end{abstract}


\section{Introduction}
Glaciers and ice sheets are machines that convert potential energy in the form of accumulated ice at high elevations into heat, either by viscous dissipation within the ice itself or by frictional dissipation at the interface between the ice and the underlying bedrock or sediment.  This latter process, hereafter referred to as `sliding', is responsible for greater than 90\% of observed surface velocity over much of Greenland, even in regions that are not particularly fast flowing \citep{Maier2019}.  Because variations in ice flow dynamics make up >50\% of contemporary ice loss in Greenland  \citep{mouginot2019forty}, correctly modeling sliding is as critical to predicting future Greenland mass loss as having reliable climate models.  Ensemble modelling of Greenland's future has shown that uncertainty in ice dynamics accounts for between 26\% and 53\% of variance in sea level rise projections over the next century \citep{aschwanden2019contribution}.

Observations \citep[e.g.,][]{Iken1986} and theoretical considerations \citep[e.g., ][]{fowler1979mathematical,weertman1964theory,lliboutry1968general} suggest that basal sliding depends on basal effective pressure. However, explicitly modelling basal effective pressure---and more generally, modeling the subglacial hydrologic system---remains among the most significant open problems in glacier dynamics. The difficulty results from a discrepancy in spatial and temporal scales between the physics driving sliding and water flux versus the scale of glaciers and ice sheets: physics at the bed occur on the order of a few meters with characteristic time scales of minutes, while relevant time scales for ice sheet evolution occur over kilometers and years. To upscale glacier hydrology to a scale relevant to the overlying ice, a variety of approximations have been proposed, including different physical phenomena thought to be morphologically relevant such as a continuum approximation of linked cavities \citep{Bueler2015}, a lattice model of conduits, or a combination thereof \citep{werder2013modeling,sommers2018shakti,hoffman2016greenland,de2014double,downs2018dynamic}. However, validating models of sliding and hydrology remains elusive, partly due to potential model misspecification, but also due to a lack of sufficient observational constraints on model parameters such as hydraulic conductivity of different components of the subglacial system, characteristic length scales of bedrock asperities, and the scaling between effective pressure and basal shear stress.

\subsection{Previous assimilation of surface velocity observations}
The above challenges are not new, and ice sheet modellers have used geophysical inversion methods \citep[e.g.,][]{parker1994geophysical} in glaciological applications to circumvent them for over two decades \citep[e.g.,][]{macayeal1993tutorial,morlighem2010spatial,gillet2012greenland,Joughin2014,Favier2014,Cornford2015}. Commonly, a linear relationship between basal shear stress and velocity is adopted, and then surface velocities are inverted for a spatially varying basal stickiness field such that the resulting surface velocities are close to observations.   This approach lumps all basal processes into one field, a frictional parameter that varies in space while ignoring temporal variability, exchanging the capability of longer-term predictive power for spatial fidelity to observations at an instant.

Several variants on this approach exist.  For example \citet{Habermann2012} performed the above procedure with a pseudo-plastic power law.  \cite{larour2014inferred} assimilated surface altimetry data into re-constructions of transient ice flow. The novelty of their approach was that surface mass balance and basal friction were determined in time as well as space, resulting in adjusted modeled surface heights and time-varying velocities that best fit existing altimetry. Such an approach allows for a better quantification of time-evolving basal and surface processes and a better understanding of the physical processes currently missing in transient ice-flow models.  Their work also demonstrated that large spatial and temporal variability is required in model characteristics such as basal friction.  However, for prognostic modeling, such approaches cannot be applied because we cannot assimilate future observations.  As such, a middle ground between purely empirical and local process modelling must be found.     

Several recent works have taken this approach.  \citet{pimentel2011numerical} used a coupled flowband model of glacier dynamics and hydrology to model the propagation of meltwater induced acceleration across a synthetic Greenland-esque domain, and established that the presence of channels can substantially reduce the sensitivity of the system to fast influxes of meltwater.  \citet{hoffman2016greenland} showed that for a 3D synthetic domain based on West Greenland, a weakly-connected drainage system helps to explain the temporal signal of velocity in the overlying ice.  The previous two studies, while not formally assimilating observations, compared their model results to observations in an effort to validate their qualitative results.  \citet{minchewd2016plastic} directly inverted surface velocities at Hofsjokull ice cap for a spatially varying basal shear stress, and in conjunction with a Coulomb friction law, inferred the distribution of effective pressure.  \cite{brinkerhoff2016inversion} used a Bayesian approach to condition a 0D model of glacier hydrology and sliding on surface velocity and terminus flux observations to infer probability distributions over unknown ice dynamics and hydrologic model parameter.  Although not coupled to a ice dynamics model, \citet{irarrazaval2019bayesian} present a Bayesian inference over the lattice model of \citet{werder2013modeling}, constraining the position and development of subglacial channels from observations of water pressure and tracer transit times.  \citet{Aschwanden2016} demonstrate that outlet glacier flow can be captured using a simple local model of subglacial hydrology, but further improvements are required in the transitional zone with speeds of 20--100\,m per year.  This disagreement between observed and simulated speeds most likely arises from inadequacies in parameterizing sliding and subglacial hydrology.  Finally and notably, \citet{koziol2018modelling} inverted velocity observations from West Greenland to determine a spatially-varying traction coefficient after attenuation by effective pressure derived from a hydrologic model.  

\subsection{Our approach}
In this work we seek to expand on previous approaches by coupling a state of the art spatially-explicit hydrology model to a 3D model of ice dynamics through a general sliding law (hereafter referred to as the \emph{high-fidelity model}), and to then infer the distribution of practically unobservable model parameters such that the ice surface velocity predicted by the model is statistically consistent with spatially explicit observations over a region in western Greenland.  Throughout the work, \emph{we assume spatially and temporally constant model parameters} so that modelling errors cannot be aliased into a large array of unconstrained basal traction values.  

It is likely that there exists substantial non-uniqueness in model parameter solutions.  Different controlling factors in the hydrology model may compensate for one another, as may parameters in the sliding law: for, example, the basal traction coefficient could be made lower if sheet conductivity is made higher, leading to a lower mean effective pressure.  In order to fully account for these tradeoffs and to honestly assess the amount of information that can be gained by looking solely at surface velocity, we adopt a Bayesian approach \citep[e.g.][]{tarantola2005inverse} in which we characterize the complete joint posterior probability distribution over the parameters, rather than point estimates.  

Inferring the joint posterior distribution is not analytically tractable, so we rely on numerical sampling via a Markov Chain Monte Carlo (MCMC) method instead.  Similar inference in a coupled hydrology-dynamics model has been done before \citep{brinkerhoff2016inversion}.  However, in previous work the model was spatially averaged in all dimensions, and thus inference was over a set of coupled ordinary differential equations.  Here, we work with a model that remains a spatially explicit and fully coupled system of partial differential equations.  As such, the model is too expensive for a naive MCMC treatment.  To skirt this issue, we create a so-called \emph{surrogate model}, which acts as a computationally efficient approximation to the expensive coupled high-fidelity model.  

To construct the surrogate, we run a 5000 member ensemble of multiphysics models through time, each with parameters drawn from a prior distribution, to produce samples of the modelled annual average velocity field.  This is computationally tractable because each of these model runs is independent, and thus can be trivially parallelized.  We reduce the dimensionality of the space of these model outputs through a principal component analysis \citep{shlens2014tutorial}, which identifies the key modes of model variability.  We refer to these modes as \emph{eigenglaciers}, and (nearly) any velocity field producible by the high fidelity model is a linear combination thereof.  To make use of this decomposition, we train an artificial neural network \citep{goodfellow2016deep} to control the coefficients of these eigenglaciers as a function of input parameter values, yielding a computationally trivial map from parameter values to a distributed velocity prediction consistent with the high fidelity model.  Unfortunately, neural networks are high variance maps, which is to say that the function is sensitive to the choice of training data.  To reduce this variance (and to smooth the relationship between parameters and predictions), we employ a Bayesian bootstrap aggregation approach \citep{clyde2001bagging,breiman1996bagging} to generate a committee of surrogate models, which are averaged to yield a prediction.    

Surrogate in hand, we use the manifold Metropolis-Adjusted Langevin Algorithm \citep[mMALA][]{girolami2011riemann} to draw a long sequence of samples from the posterior probability distribution of the model parameters.  mMALA utilizes both gradient and Hessian information that are easily computed from the surrogate to efficiently explore the posterior distribution.  Because the surrogate model itself is based on a finite sample of a random function, we use a second Bayesian bootstrap procedure to integrate over the surrogate's random predictions, effectively accounting for model error in posterior inference \citep{huggins2019using} induced by using the surrogate (rather than the high fidelity model) for inference.

We find that high fidelity model is able to reproduce many of the salient features of the observed annual average surface velocity field for a terrestrially terminating subset of southwestern Greenland, with the model explaining on average around 60\% of the variance in observations.  As expected, we find significant correlations in the posterior distribution of model parameters.  However, we also find that surface velocity observations provide substantial constraints on most model parameters.  To ensure that the distribution inferred using the surrogate is still reasonable given the high fidelity model, we select a handful of samples from the posterior distribution, feed them back into the high fidelity model, and show that the resulting predictive distribution remains consistent with observations.  The process described above is applicable to the broad class of problems in which we would like to perform Bayesian inference over a limited number of parameters given an expensive deterministic model.  As such, we first describe the method in detail, then discuss its glaciological implications.  

\section{Study Area}
\label{sec:study_area}
\begin{figure*}
    \centering
    \includegraphics[width=0.9\linewidth]{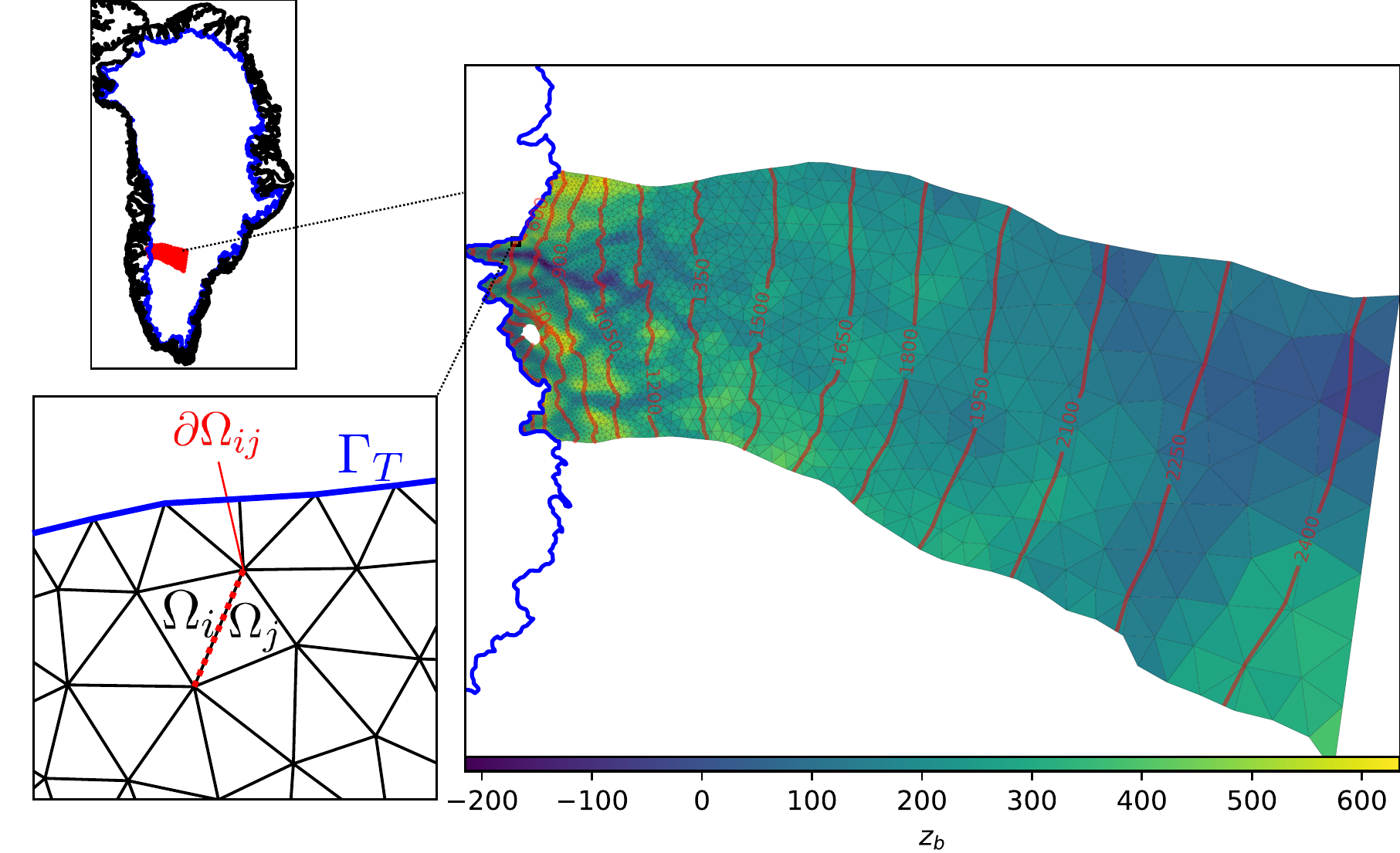}
    \caption{Study area, with location of domain in Greenland (upper left), detailed modelling domain with the computational mesh overlain with bedrock elevation and surface contours (right), and closeup of mesh with domains used in modelling labelled (lower left, see text).}
    \label{fig:study_area}
\end{figure*}

We focus our study on the region of western Greenland centered around Russell Glacier (Fig. \ref{fig:study_area}). The domain runs from the ice margin to the ice divide, covering an area of approximately 36,000km$^2$.  This region was selected because it strikes a balance between being simple and being representative: all glacier termini are terrestrial, which means that the effects of calving can be neglected in this work, surface slopes are modest, and surface meltwater runoff rates are neither extreme nor negligible, yet there is still substantial spatial variability in glacier speed even near the margin, from a maximum of 150\,m/a over the deep trench at Isunnguata Sermia, to less than 30 m/a just 20 km to the north.  

Additionally, this region of Greenland has long been a hotspot for observations due to its proximity to the town of Kangerlussuaq.  The bed is well-constrained by Operation IceBridge flightlines, and throughout this work, we use the basal topography of BedMachine V2 \citep{morlighem2017bedmachine}.  We force the model with surface meltwater runoff computed with HIRHAM \citep{hirham2017}, averaged by month between 1992 and 2015.  As such, our forcing is time-varying but periodic with a period of one year.  When comparing modelled to observed velocities (henceforth called $\mathbf{u}_{obs}$), we use as our observation the inSAR derived annual average velocity fields of \citet{joughin2017complete}, further averaged over the years 2014 through 2018.  

\section{Numerical Models}

    \begin{longtable}{c|c|c|l}
        Symbol & Value & Units & Description \\ \hline
        $A$ & $10^{-16}$ & Pa$^{-n}$ a$^{-1}$ & Ice softness \\
        $\beta^2$ & & Pa$^{1-p}$ m$^{-q}$ a$^{-q}$ & Traction coefficient \\
        $e_v$  & & & Englacial porosity \\
        $\dot{\epsilon}$ & & a$^{-1}$ & Strain rate tensor \\
        $\dot{\epsilon}_0$ & $10^{9}$ & a$^{-1}$& Strain rate regularization \\
        $\dot{\epsilon}_{II}$ & & a$^{-1}$&  Second invariant of strain rate tensor \\
        $g$ & 9.81 & m s$^{-2}$ & Gravitational acceleration \\
        $\gamma$ & & Pa & Scaled traction coefficient \\
        $\Gamma_{z_b}$ & &  & Basal boundary \\
        $\Gamma_L$ & &  & Non-terminus lateral boundary \\
        $\Gamma_{z_s}$ & &  & Surface boundary \\
        $\Gamma_T$ & &  & Terminal boundary \\
        $h$ & & m & Average cavity thickness \\
        $\bar{h}_r$ & & m & Average bedrock bump size \\
        $H$ & & m & Ice thickness \\
        $\eta$ & & Pa a & Ice viscosity \\
        $L$ & 3.35$\times 10^5$ & J kg$^{-1}$ & Latent heat of fusion \\
        $k_c$ & & m$^{2 - 2\alpha + \beta}$ a$^{-1}$ Pa$^{1-\beta}$ & Channel conductivity \\
        $k_s$ & & m$^{1 - \alpha + \beta}$ a$^{-1}$ Pa$^{1-\beta}$ & Sheet conductivity \\
        $\dot{m}$ & & m a$^{-1}$ & Specific meltwater \\
        $m_c$ & & m$^2$ a$^{-1}$ & Channel-cavity meltwater exchange \\
        $n$ & 3 & & Glen's flow law exponent\\
        $n_p$ & & & Number of points in FEM mesh\\
        $\mathbf{n}$ & &  & Normal vector \\
        $N$ & & Pa & Effective pressure \\
        $p$ & & & Sliding law pressure exponent \\
        $P_0$ & & Pa & Ice overburden pressure \\
        $P_w$ & & Pa & Water pressure \\
        $\Pi$ & & & Finite element basis function \\
        $q$ & & & Sliding law velocity exponent \\
        $\mathbf{q}$ & & m$^2$ a$^{-1}$ & Cavity flux \\
        $Q$ & & m$^3$ a$^{-1}$ & Channel discharge \\
        $r$ & & & Ratio of asperity height to spacing \\
        $\rho_i$ & 917 & kg m$^{-3}$ & Ice density \\
        $\rho_w$ & 1000 & kg m$^{-3}$ & Freshwater density \\
        $S$ & & m$^2$ & Channel size \\
        Scale($N$) & $10^6$ & Pa & Effective pressure scale \\
        Scale($\mathbf{u}$) & 50 & m a$^{-1}$ & Velocity scale \\
        $\sigma_h^2$ & & m & logarithmic std. dev. of bed asperity size \\
        $\varsigma$ & & & Thickness-scaled vertical coordinate \\
        $\tau'$ & & Pa & Hydrostatic deviatoric stress tensor \\
        $\mathbf{\tau}_d$ & & Pa & Driving stress \\
        $\mathbf{u}$ & & m a$^{-1}$ & Horizontal velocity vector \\
        $\bar{\mathbf{u}}$ & & m a$^{-1}$ & Vertically-averaged velocity vector \\
        $\mathbf{u}_d$ & & m a$^{-1}$ & Shear velocity \\
        $\phi$ & & Pa & Hydraulic potential \\
        $\xi$ & & & Lagrange basis function \\
        $\Xi$ & & J m$^{-1}$ a$^{-1}$ & Dissipative heating \\
        $\Psi$ & & J m$^{-1}$ a$^{-1}$ & Pressure heating \\
        $z_b$ & & m & Bed elevation \\
        $z_s$ & & m & Surface elevation \\ 
        $\Omega$ & & & 3D ice domain \\
        $\delta \Omega_{ij}$ & & & Boundary between subdomains $i$ and $j$ \\
        $\bar{\Omega}$ & & & Horizontal extent of ice \\
    \caption{Symbols used in defining the high-fidelity model.}
    \label{tab:hifi_symbols}
    \end{longtable}

We simulate surface speeds using a coupled model of ice flow and subglacial water flow.   
\subsection{Ice Dynamics}
\subsubsection{Viscous flow}

The flow of the ice sheet over a volume $\Omega$ is modelled as a low Reynolds number fluid using a hydrostatic approximation to Stokes' equations \citep{pattyn2003new}
\begin{align}
    \nabla \cdot \tau' = \rho_i g \nabla z_s,
    \label{eq:stokes}
\end{align}
where 
\begin{align}
    \tau' = \begin{bmatrix} 2\tau_{xx} + \tau_{yy} & \tau_{xy} & \tau_{xz} \\
                            \tau_{xy} & \tau_{xx} + 2\tau_{yy} & \tau_{yz}
    \end{bmatrix}.
\end{align}
$z_S$ is the glacier surface elevation, $\rho_i$ is ice density, $g$ the gravitational acceleration, and $\tau_{ij}$ is a component of the deviatoric stress tensor given by
\begin{equation}
\tau_{ij} = 2\eta \dot{\epsilon}_{ij},
\end{equation}
with $\dot{\epsilon}$ the symmetrized strain rate tensor.  The viscosity 
\begin{equation}
\eta = \frac{A}{2}^{-\frac{1}{n}}(\dot{\epsilon}_{II} + \dot{\epsilon}_0)^{1-\frac{1}{n}}
\end{equation}
is dependent on the second invariant of the strain rate tensor $\dot{\epsilon}_{II}$.  Note that we make an isothermal approximation, and take the ice softness parameter $A$ to be a constant.  The exponent in Glen's flow law $n$ is taken to be three.
\subsubsection{Boundary Conditions}
At the ice surface $\Gamma_{z_s}$ and terminal margin $\Gamma_{T}$, we impose a no-stress boundary condition
\begin{align}
    \tau' \cdot \mathbf{n} = \mathbf{0},
\end{align}
where $\mathbf{n}$ is the outward pointing normal vector, and $\mathbf{0}$ is the zero vector.  At the remaining lateral boundaries $\Gamma_L$, we do not impose a boundary condition (i.e. we allow a free flux of momentum across the boundary).  

At the basal boundary $\Gamma_{z_B}$ we impose the sliding law
\begin{equation}
\label{eq:sliding_law}
\tau'\cdot \mathbf{n} = -\beta^2 N^p \|\mathbf{u}\|_2^{q-1} \mathbf{u},
\end{equation}
with $\beta^2$ the basal traction coefficient and $\mathbf{u}$ the ice velocity.  We note that this sliding law has some theoretical \citep{fowler1987sliding} and empirical \citep{budd1979empirical,bindschadler1983importance} support, but does not satisfy Iken's bound \citep{iken1981effect}.  As such there are alternative sliding laws that may be preferable \citep[e.g][]{schoof2005effect}.  However, we defer a detailed comparison of different sliding laws and condition this work on Eq.~\ref{eq:sliding_law} being a reasonable (and numerically stable) approximation to the true subglacial process.

The effective pressure $N$ is given by the ice overburden pressure $P_0$ less the water pressure $P_w$
$$
N = P_0 - P_w.
$$ 
The exponents $p$ and $q$ control the non-linear response of basal shear stress to the effective pressure and velocity (respectively).  We note several limiting cases of this sliding law: when $p=q=1$, we recover the linear Budd law \citep{budd1979empirical}.  When $p=0$, we get the pressure independent Weertman law \citep{weertman1957sliding}.  In the limit $q\rightarrow\infty$, we recover a perfectly plastic model of basal stress \citep[e.g][]{kamb1991rheological}.

In practice, we use a reparameterized version of Eq.~\ref{eq:sliding_law}
\begin{equation}
    \tau_b = \gamma^2 \hat{N}^p \|\hat{\mathbf{u}}\|^{q-1} \hat{\mathbf{u}},
\end{equation}
where $\frac{N}{\mathrm{Scale}(N)} = \hat{N}$ is the effective pressure non-dimensionalized by a constant factor (here we use the ice overburden averaged over the model domain), and $\frac{\mathbf{u}}{\mathrm{Scale}(\mathbf{u})} = \hat{\mathbf{u}}$ is similar, with the characteristic scale of $\mathbf{u}$ taken to be 50 m/a.  Thus, the resulting relationship between $\gamma^2$ (which has units of stress) and $\beta^2$ is
\begin{equation}
\beta^2 = \frac{\gamma^2}{\mathrm{Scale}(N)^p \mathrm{Scale}(\mathbf{u})^q}.
\end{equation}
This transformation is helpful because the power law terms on the right hand side of Eq.~\ref{eq:sliding_law} can vary by several orders of magnitude, thus requiring that $\beta^2$ does the same in order to maintain a given characteristic surface velocity.  The $\gamma^2$ parameterization circumvents this scale issue.  We take $\gamma^2$, $p$, and $q$ to be unknown but spatially and temporally constant.

\subsubsection{Discretization}
We discretize the momentum equations using a mixed finite element method.  Introducing a terrain-following $\varsigma$-coordinate
\begin{equation}
\varsigma = \frac{z_s-z}{H},
\end{equation}
where $z_s$ is the upper ice surface, $H$ is ice thickness and $z$ the vertical coordinate, we decompose the domain as $\Omega = \bar{\Omega} \times [0,1]$.  Introducing a test function $\Pi(x,y,\varsigma)$, multiplying it by Eq.~\ref{eq:stokes}, and integrating over the domain, we obtain the following variational formulation: find $\mathbf{u} \in U$, such that
\begin{align}
    0 &= \int_{\bar{\Omega}} \int_1^0 (\bar{\nabla} \Pi + \partial_\varsigma \Pi \bar{\nabla} \varsigma) \cdot \tau' \,H \;\mathrm{d}\varsigma \;\mathrm{d}\Omega \nonumber \\
      & - \int_{\Gamma_l} \int_1^0 \Pi \cdot \tau' \cdot \mathbf{n} \varsigma\; \mathrm{d}\Gamma \nonumber \\
      & - \int_{\bar{\Omega}} \int_1^0 \Pi \cdot \tau_d \;\mathrm{d}\varsigma\;\mathrm{d}\Omega \nonumber \\
      & + \int_{\bar{\Omega}} \Pi \cdot \beta^2 N^p \|\mathbf{u}\|_2^{p-1} \mathbf{u} \,\mathrm{d}\Omega |_{\varsigma=1}, \nonumber \\
      & \forall\Pi\in V,
\end{align}
with $U,V\in W^{1,2}(\Omega)$.  To discretize the weak form, we restrict $\Pi$ to a finite subset of $V$:
\begin{equation}
    \Pi \in \hat{V} \subset V,
\end{equation}
where 
\begin{equation}
\hat{V} = V_{\bar{\Omega}} \otimes V_{\bar{\Omega}} \otimes V_0 \otimes V_0
\end{equation}
is a tensor product of function spaces defined over $\bar{\Omega}$ and $[0,1]$, respectively.  For $V_{\bar{\Omega}}$, we use the standard first order Lagrange (i.e. local piecewise linear) basis $\left\{\xi_i\right\}_{i=1}^{n_p}$, where $n_p$ is the number of grid points in a mesh defined on $\bar{\Omega}$ \citep{zienkiewicz2005finite}.  For $V_0$, we utilize the basis set 
\begin{equation}
    \left\{ \psi_1 = 1, \psi_2 = \frac{1}{n+1} [(n+2) \varsigma^{n+1} - 1] \right\}.
\end{equation}
Using the standard Galerkin approximation $\hat{U}=\hat{V}$, we introduce the ansatz solution
\begin{equation}
\mathbf{u}(x,y,\varsigma) = \sum_{i\in n}\left[\bar{\mathbf{u}}_i + \mathbf{u}_{d,i} \frac{1}{n+1} [(n+2)\varsigma^{n+1} - 1] \right]\xi_i(x,y),
\end{equation}
where $\bar{\mathbf{u}}$ is the vertically averaged velocity, and $\mathbf{u}_d$ is the deviation from that average induced by vertical shearing.  The above expression implies that the solution in the vertical dimension is a linear combination of a constant (i.e. the shallow-shelf approximation) and a polynomial of order $n+1$, which corresponds to the analytical solution of the isothermal shallow ice approximation.  As such, this discretization scheme allows for the exact recovery of both shallow ice and shallow shelf solutions in the appropriate asymptotic regimes, while not requiring the formation of a full three dimensional mesh (the $\varsigma$ dimension always has one layer, ranging over $\varsigma\in[0,1]$).  Intercomparison has shown that approximate solutions produced by this method agree well with more expensive three-dimensional discretizations of the hydrostatic Stokes' equations \citep{brinkerhoff2015dynamics}.    

\subsection{Hydrologic model}
In order to predict the effective pressure $N$ on which the sliding law depends, we couple the above ice dynamics model to a hydrologic model that simulates the evolution of the subglacial and englacial storage via fluxes of liquid water through an inefficient linked cavity system and an efficient linked channel system.  This model closely follows the model GlaDS \citep{werder2013modeling}, with some alterations in boundary conditions, discretization, and opening rate parameterization.  

Over a disjoint subdomain $\bar\Omega_i \subset \bar\Omega,\; \bigcup_{i\in\mathcal{T}} \bar\Omega_i=\bar{\Omega}$, where $\mathcal{T}$ is the set of triangles in the finite element mesh, the hydraulic potential $\phi$ evolves according to the parabolic equation
\begin{equation}
    \frac{e_v}{\rho_w g} \frac{\partial \phi}{\partial t} + \nabla \cdot \mathbf{q} - \mathcal{C} + \mathcal{O} = m,
    \label{eq:hydraulic_potential}
\end{equation}
where $\phi$ is the hydraulic potential, $\rho_w$ the density of water, $\mathbf{q}$ the horizontal flux, $\mathcal{C}$ the rate at which the cavity system closes (pushing water into the englacial system), $\mathcal{O}$ the rate at which it opens, and $m$ is the recharge rate (either from the surface, basal melt, or groundwater).  The horizontal flux is given by the Darcy-Weisbach relation
\begin{equation}
    \mathbf{q} = -k_s h^{\alpha_s} \|\nabla \phi\|_2^{\beta_s-2} \nabla \phi,
\end{equation}
a non-linear function of the hydraulic potential, characteristic cavity height $h$, bulk conductivity $k_s$, and turbulent exponents $\alpha_s$ and $\beta_s$.  

The average subglacial cavity height $h$ evolves according to
\begin{equation}
\frac{\partial h}{\partial t} = \mathcal{O} - \mathcal{C}.
  \label{eq:cavity_evolution}
\end{equation}
Here we model the subgrid-scale glacier bed as self-similar, with bedrock asperity heights modeled with a log-normal distribution:
\begin{equation}
    \log h_r \sim \mathcal{N}(\log \bar{h}_r, \sigma_h^2),
\end{equation}
and a characteristic ratio $r$ of asperity height to spacing.  Thus, the opening rate is given by 
\begin{equation}
    \mathcal{O} = \int_{0}^\infty \mathrm{Max}\left(\|\mathbf{u}(\varsigma=1)\|_2 r (1 - \frac{h}{h_r}),0\right) \; P(h_r)\; \mathrm{d}h_r,
\end{equation}
where we use $P(\cdot)$ to denote a probability density function.  For $\sigma_h^2=0$, this expression is equivalent to the standard opening rate used in previous works \citep[e.g.][]{werder2013modeling}, albeit reparameterized.  However, this implies that once the cavity size reaches $h_r$, then the opening rate becomes zero: for a glacier moving increasingly quickly due to a high water pressure, there is no mechanism for subglacial storage capacity to increase.  For $\sigma_h^2>0$, our formulation regularizes the opening rate such that there is `always a bigger bump,' but with a diminishing effect away from the modal bump size.  Here, we make the somewhat arbitrary choice that $\sigma_h^2=1$ and take $\bar{h_r}$ to be a tunable parameter.

The cavity closing rate is given by 
\begin{equation}
    \mathcal{C} = \frac{2}{n^n} A h |N|^{n-1} N.
\end{equation}

Over a domain edge $\partial \Omega_{ij}$ (the edge falling between subdomains $\bar\Omega_i$ and $\bar\Omega_j$, mass conservation implies that 
\begin{equation}
    \frac{\partial S}{\partial t} + \frac{\partial Q}{\partial s} = \frac{\Xi - \Pi}{\rho_w L} + m_c,
    \label{eq:channel_mass_cons}
\end{equation}
with $S$ the size of a channel occurring along that edge, $\Xi$ the opening rate due to turbulent dissipation, $\Pi$ the rate of sensible heat changes due to pressure change, and $m_c$ the exchange of water with adjacent domains.  The channel discharge $Q$ is given by another Darcy-Weisbach relation
\begin{equation}
    Q = -k_c S^\alpha \|\frac{\partial \phi}{\partial s} \|^{\beta-2} \frac{\partial \phi}{\partial s},
\end{equation}
where $k_c$ is a bulk conductivity for the efficient channelized system.  The channel size evolves according to 
\begin{equation}
    \frac{\partial S}{\partial t} = \frac{\Xi - \Pi}{\rho_i L} - \mathcal{C}_c,
    \label{eq:channel_evol}
\end{equation}
with channel closing rate
\begin{equation}
    \mathcal{C}_c = \frac{2}{n^n} A S |N|^{n-1} N.
\end{equation}
Substitution of Eq.\ref{eq:channel_evol} into Eq.\ref{eq:channel_mass_cons} leads to an elliptic equation 
\begin{equation}
    \frac{\partial Q}{\partial s} = \frac{\Xi - \Pi}{L} \left(\frac{1}{\rho_i} - \frac{1}{\rho_w}\right) + m_c. 
    \label{eq:elliptic_channel}
\end{equation}
The exchange term with the surrounding sheet is given by 
\begin{equation}
    [\mathbf{q}\cdot \mathbf{n}]_+ + [\mathbf{q} \cdot \mathbf{n}]_- = m_c,
\end{equation}
which states that flux into (or out of) a channel is defined implicitly by the flux balance between the two adjacent sheets.
\subsubsection{Boundary Conditions}
We impose a no-flux boundary condition across boundaries $\Gamma_T \cup \Gamma_L$ in both the sheet and conduit model:
\begin{align}
    \mathbf{q}\cdot \mathbf{n} &= 0 \\
    Q = 0 
\end{align}
At first glance, this seems to be a strange choice: how then, does water exit the domain?  To account for this, we impose the condition that whenever $\phi>\phi_S$, where $\phi_{z_s}=\rho_w g z_s$ is the surface potential, any excess water immediately runs off.  Because the margins are thin, and the flux across the lateral boundary is zero, the hydraulic head there quickly rises above the level of the ice surface, and the excess water runs off.  This heuristic is necessary to avoid the numerically challenging case when potential gradients would imply an influx boundary condition.  With a free flux boundary, the model would produce an artificial influx of water from outside the domain in order to keep channels filled, which is particularly problematic in steep topography.  Most of the time, the chosen inequality condition has the practical effect of setting the hydraulic potential at the terminus to atmospheric pressure.  

In addition to this condition, we also enforce the condition that channels do not form at the margins (i.e. $S=0$ on $\Gamma_T \bigcup \Gamma_L$).  At the terminus, this ensures that there are no channels with unbounded growth perpendicular to the terminus, and also to ensure that lateral boundaries (where $H>0$) are not used as preferential flow paths.  

\subsubsection{Discretization}
We seek to solve Eq.~\ref{eq:hydraulic_potential} on each subdomain $\bar\Omega_j$ and Eq.\ref{eq:elliptic_channel} on each subdomain boundary $\Gamma_{ij}$.  To discretize, we multiply both by the same test function $\theta$ and integrate by parts, leading to the variational problem: find $\phi\in \Phi$ such that
\begin{align}
    0 =\sum_j \int_{\bar\Omega_j} \theta \frac{e_v}{\rho_w g}\frac{\partial \phi}{\partial t} - \nabla \theta \cdot \mathbf{q} + \theta(\mathcal{C} - \mathcal{O} - m) \,\mathrm{d}\Omega \\
    + \sum_j \sum_{i<j} \int_{\Gamma_{ij}} -\frac{\partial \theta}{\partial S} Q + \theta\left(\frac{\Xi - \Pi}{L}\left(\frac{1}{\rho_i} - \frac{1}{\rho_w}\right) - \mathcal{C}_c\right)\;\mathrm{d}\Gamma\\
    \forall \theta \in \Theta,
\end{align}
where $\Phi,\Theta \in W^{1,2}(\bar\Omega)$.  We have used natural boundary conditions, continuity between channel segments, and continuity between the sheet and edges to cancel boundary terms.  To discretize this equation, we restrict $\hat{\Phi}\subset \Phi,\hat{\Theta} \subset \Theta$ to function spaces defined by the standard linear Lagrange basis.  

Although Eqs.~\ref{eq:cavity_evolution} and \ref{eq:channel_evol} are ordinary differential equations, it is convenient to put them in a variational form: find $h \in Z,S\in \Sigma$ such that
\begin{align}
    0 &= \sum_j \int_{\Omega_j }\left[\frac{\partial h}{\partial t} - \mathcal{O} + \mathcal{C}\right]w \mathrm{d}\Omega \nonumber \\
    & + \sum_i \sum_{j<i} \int_{\Gamma_{ij}} \left[\frac{\partial S}{\partial t} - \frac{\Xi - \Pi}{\rho_i L} + \mathrm{C}_c\right]v \mathrm{d}\Gamma, \nonumber \\
    \forall w\in Z , v\in\Sigma,
    \end{align}
with $Z\in L^2(\bar\Omega)$, $\Sigma \in L^2(\Gamma)$.  We restrict these functions spaces to a constant basis over each subdomain (i.e. Order-zero discontinuous Galerkin over both mesh elements and edges).  

\subsection{Numerical Solution}
We use the finite element software FENiCS  \citep{logg2012automated} to compile all of the variational problems described above.  We solve the problems over an isotropic computational mesh with variable resolution, from approximately 250m diameter elements near the margins, to approximately 1km near the ice divide.  The mesh was created using a Delaunay Triangulation routine in the package gmsh \citep{geuzaine2009gmsh}.  

We use the implicit Euler method \citep{butcher2016numerical}  to discretize all time steps.  While less accurate, we have found that the implicit Euler method leads to substantially improved stability in the non-linear cavity and conduit equations.  We deal with the integral in $\mathcal{O}$ using Gauss-Legendre numerical quadrature of order seven \citep{milne1972handbook}.

Because the system of equations are non-linear and strongly coupled, we perform Newton's method on a single residual encompassing all seven equations \emph{simultaneously}, using a Jacobian inferred from an automated symbolic computation of the Gateaux derivative.  Note that this implies that we must solve a large non-linear system at each time step.  Because of the poor conditioning of the problem, we have found direct solution of the linear system of equations for each Newton update is required.  To this end, we use MUMPS, which is implemented in PETSc \citep{balay2017petsc}.  

We employ an adaptive time-stepping procedure that ensures convergence: the time step is slowly increased until Newton's method fails to produce a residual with a specified tolerance within a certain number of iterations, at which point the time step is reduced by half and the solver tries again until convergence is achieved, after which time-stepping proceeds.  

\section{Surrogate Model}
    \begin{longtable}{c|l}
        Symbol & Description \\ \hline
        $a$ & MCMC acceptance probability \\
        $\hat{\mathbf{a}}_l$ & Output of linear transform \\ 
        $\mathbf{a}_l$ & Output of layer normalization \\
        $\alpha_l$ & Layer normalization scaling \\
        $\alpha$ & Prior parameter \\
        $\mathbf{b}_l$ & Trainable bias vector \\
        $\beta_l$ & Layer normalization offset \\
        $\beta$ & Prior parameter \\
        $\mathrm{Bound}_L$ & Parameter lower bound \\
        $\mathrm{Bound}_U$ & Parameter upper bound \\
        $c$ & Number of retained eigenglaciers \\
        $d(x,x')$ & Distance \\
        $\mathbf{d}$ & Data vector \\
        $\Delta$ & MCMC step size \\
        $f$ & Fraction of explained variance \\
        $\mathcal{F}$ & High-fidelity model \\
        $\mathcal{G}$ & Surrogate model \\
        $\mathbf{h}_l$ & Residual sum \\
        $\hat{H}$ & Approximate Hessian \\
        $k$ & Parameter vector length \\
        $K$ & Number of observations per subdomain matrix \\
        $l$ & Length scale of data correlation \\
        $L$ & Number of ANN blocks \\
        $\mathbf{m}$ & Vector of model parameters \\
        $P_{em}(\mathbf{m})$ & Evaluation sampling distribution \\ 
        $Q(\cdot|\cdot)$ & MCMC proposal function \\ 
        $r(x)$ & Data residual function\\
        $R$ & Dropout matrix \\
        $\mathbf{r}$ & Residual vector \\
        $\rho_d$ & Data density \\
        $s$ & Explained variance threshold \\
        $\sigma(x,x')$ & Covariance function \\
        $\sigma_{obs}$ & Data white noise std. \\
        $\sigma_{cor}$ & Data correlated noise std. \\
        $\mathcal{S}$ & Model empirical covariance \\
        $\hat{\Sigma}$ & Data covariance matrix \\
        $\Sigma$ & Area-scaled data covariance matrix \\
        $V$ & Matrix of ensemble eigenvectors \\
        $\lambda$ & Eigenglacier coefficients \\
        $\Lambda$ & Diagonal matrix of ensemble eigenvalues \\
        $\theta$ & Surrogate model trainable parameters \\
        $W_l$ & Trainable weight matrix\\
        $\hat{\mathbf{z}}_l$ & Output of activation \\
        $\mathbf{z}$ & Output of dropout \\ 
        $\omega_d$ & Vector of bootstrap weights for surrogate training \\
        $\omega_e$ & Vector of bootstrap weights for aggregation \\

    \caption{Symbols used in defining the surrogate model and MCMC sampling.}
    \label{tab:surrogate_symbols}
\end{longtable}
The solution of the coupled model defined above defines a function $\mathcal{F} : \mathbb{R}_+^k \rightarrow \mathbb{R}^{n_p}$ that maps from a parameter vector   
\begin{equation}
    \mathbf{m} = [k_s,k_c,\bar{h}_r,r,\gamma^2,p,q,e_v]^T.
\end{equation}
of length $k=8$ to a vector of annually-averaged surface speeds defined at each point on the computational mesh 
$$
\mathcal{F}(\mathbf{m})=\frac{1}{t_1-t_0}\int_{t_0}^{t_1} \|\mathbf{u}(t;\mathbf{m})|_{z=z_s}\|_2  \mathrm{d}t,
$$
where $t_0=15$ and $t_1=20$, i.e. the result of running the high-fidelity model with time-varying meltwater forcing for 20 years given parameters $\mathbf{m}_i$, computing the speed at the surface, and taking its average over the last five years to ensure that the model has reached dynamic equilibrium.  The evaluation of $\mathcal{F}$ is computationally expensive.  However, we anticipate needing to evaluate it many times in order to approximate parameter uncertainty through, for example, an MCMC sampling scheme, which cannot be easily parallelized.  We therefore seek to create a function $\mathcal{G} : \mathbf{R}_+^k \rightarrow \mathbb{R}^{n_p}$ that yields approximately the same map as $\mathcal{F}$, but at a substantially lower cost.

A variety of mechanisms may be used to construct such an approximation, here called the \emph{surrogate model}.  To construct the surrogate, we take a machine learning approach, in which we create a large (but finite) set of model input and output pairs $D = \{(\mathbf{m}_i,\mathcal{F}(\mathbf{m}_i))\}$.   We then use these input-output pairs as training examples over which to optimize the parameters of a highly flexible function approximator, in this case an artificial neural network.  We note that each sample is independent, and thus the evaluation of the high-fidelity model for each ensemble member can be performed with perfect parallelism. 

\subsection{Large Ensemble}

\begin{table}[]
    \centering
    \begin{tabular}{c|c|c}
        Parameter & Lower bound & Upper bound \\ \hline
        $k_s$ & $10^{-4}$ & $10^0$ \\
        $k_c$ & $10^{-4}$ & $10^0$ \\
        $\bar{h}_r$ & $10^{-3}$ & $10^1$ \\
        $r$ & $10^{-2}$ & $10^1$ \\
        $\gamma^2$ & $10^5$ & $10^7$ \\
        $p$ & $10^{-1}$ & $1.2$ \\
        $q$ & $10^{-1}$ & $1.2$ \\
        $e_v$ & $10^{-4}$ & $10^{-2}$ 
    \end{tabular}
    \caption{Upper and lower bounds for both the log-uniform distribution used to generate surrogate training examples, as well as the log-beta prior distribution.}
    \label{tab:bounds}
\end{table}
In order to construct the training data for $\mathcal{G}$, we must select the values $\mathbf{m}_i$ over which $\mathcal{F}$ should be evaluated.  Because all values in $\mathbf{m}$ are positive, yet we do not wish to bias the dataset towards certain regions of the plausible parameter set over others, we choose to draw $\mathbf{m}$ from a log-uniform distribution with lower and upper bounds $\mathbf{b}_L$ and $\mathbf{b}_U$:
\begin{equation}
    \log_{10}(\mathbf{m}) \sim \mathcal{U}(\mathrm{Bound}_L,\mathrm{Bound}_U).
\end{equation}
We refer to this distribution as $P_{em}(\mathbf{m})$.  The specific values of the bounds are given in Table~\ref{tab:bounds}, but in general, parameters vary a few orders of magnitude in either direction from values commonly found in the literature.  Note that this distribution is \emph{not} the prior distribution that we will use for Bayesian inference later on.  Rather, it is an extremal bound on what we believe viable parameter values to be.  However, the support for the distributions is the same, ensuring that the surrogate model is not allowed to extrapolate.  

One viable strategy for obtaining training examples would be to simply draw random samples from $P_{em}(\mathbf{m})$, and evalute the high-fidelity model there.  However, because we would like to ensure that there is a sample ``nearby'' all locations in the feasible parameter space, we instead generate the samples using the quasi-random Sobol sequence \citep{sobol2011construction}, which ensures that the parameter space is optimally filled (the sequence is constructed such that the sum of a function evaluated at these samples converges to the associated integral over the domain as quickly as possible).  While the Sobol sequence is defined over the $k$-dimensional unit hypercube, we transform it into a quasi-random sequence in the space of $P_{em}(\mathbf{m})$ using the percent point function.

With this distribution of parameters in hand, we evaluate $\mathcal{F}$ on each sample $\mathbf{m}_i$.  Using 48 cores, this process took approximately 4 days for 5000 samples.  Note that some parameter combinations never converged, in particular cases where $\gamma^2$ was too low and the resulting velocity fields were many orders of magnitude higher than observed.  We discarded those samples and did not use them in subsequent model training.  

\subsection{Surrogate Architecture}
\subsubsection{Dimensionality Reduction}
\begin{figure*}
    \centering
    \includegraphics[width=0.8\linewidth]{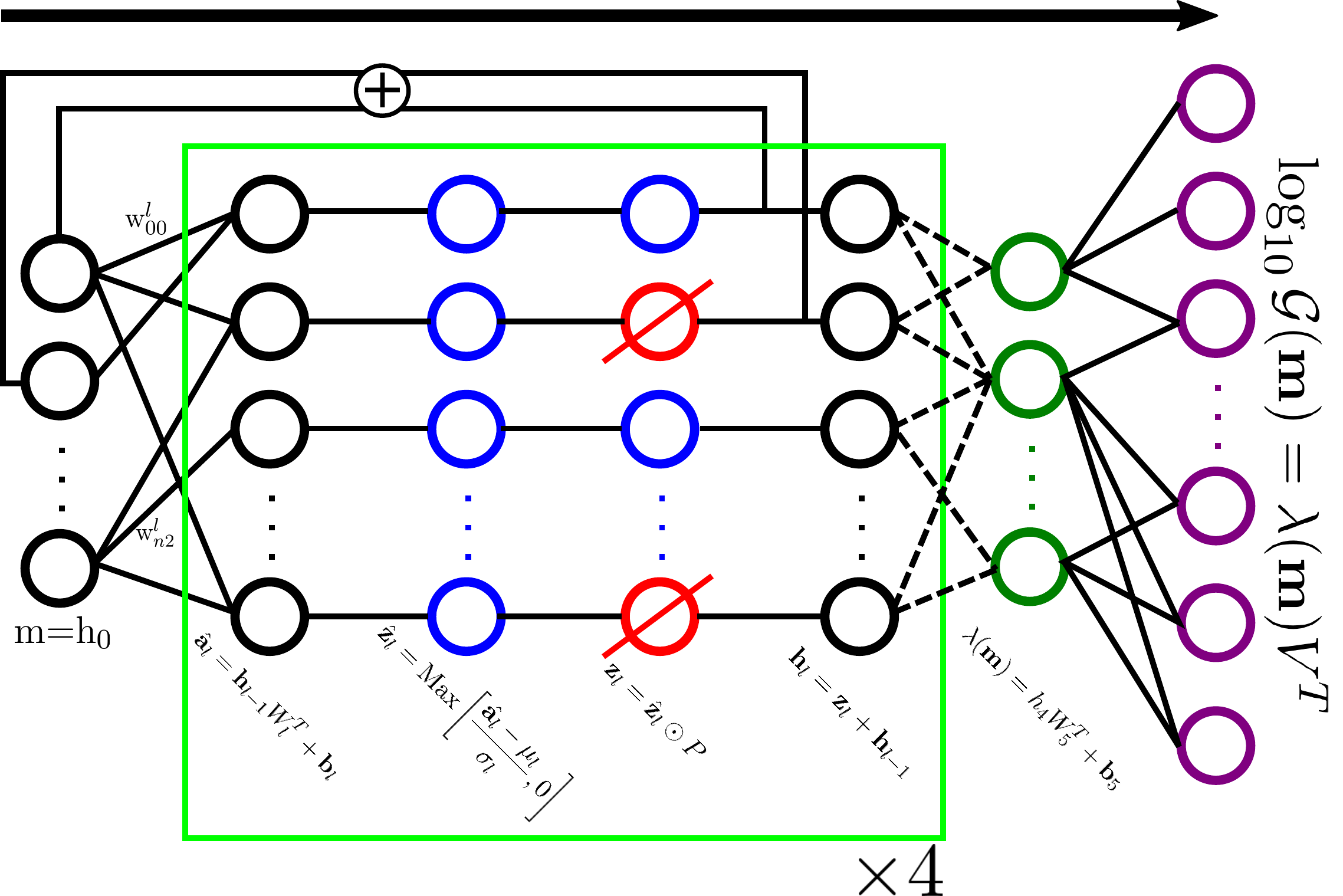}
    \caption{Architecture of the neural network used as a surrogate model in this work, consisting of 4 repetitions of linear transformation, layer normalization, dropout, and residual connection, followed by projection into the velocity field space through linear combination of basis functions computed via principal components analysis.}
    \label{fig:schematic}
\end{figure*}
We construct the surrogate model $\mathcal{G}$ in two stages.  In the first stage, we perform a principal component analysis \citep[PCA,][]{shlens2014tutorial} to extract a limited set of basis functions that can be combined in linear combination such that they explain a maximal fraction of the variability in the ensemble.  Specifically, we compute the eigendecomposition
\begin{equation}
\mathcal{S} = V \Lambda V^T,
\end{equation}
where $\Lambda$ is a diagonal matrix of eigenvalues and the columns of $V$ the eigenvectors of the empirical covariance matrix of $\log_{10} U$
\begin{equation}
    \mathcal{S} = \sum_{i=1}^m \omega_{d,i} \left[\log_{10} \mathcal{F}(\mathbf{m}_i) - \log_{10} \bar{\mathcal{F}} \right]^2,
\end{equation}
with $\omega_d$ a vector of weights such that $\sum_{i=1}^m \omega_{d,i} = 1$ and 
\begin{equation}
    \log_{10} \bar{\mathcal{F}} = \sum_{i=1}^m \omega_{d,i} \log_{10} \mathcal{F}(\mathbf{m}_i).
\end{equation}
We work with log-velocities due to the large variability in the magnitude of fields that are produced by the high fidelity model.

\begin{figure}
    \centering
    \includegraphics[width=0.6\linewidth]{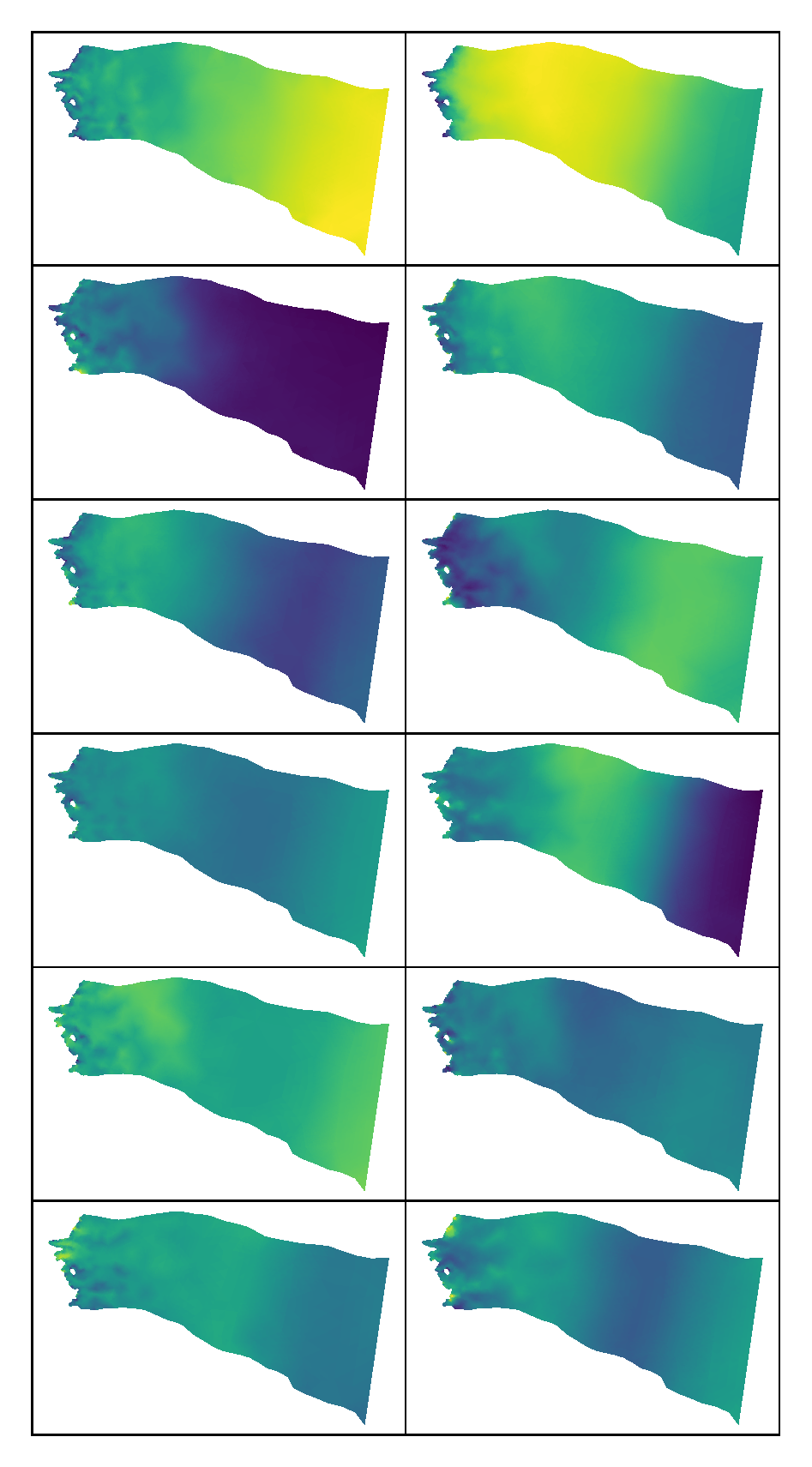}
    \caption{First 12 basis functions in order of explained variance for one of 50 bootstrap-sampled ensemble members.}
    \label{fig:basis_functions}
\end{figure}

The columns of $V$ are an optimal basis for describing the variability in the velocities contained in the model ensemble.  They represent dominant model modes (Fig.~\ref{fig:basis_functions}) (in the climate literature, these are often called empirical orthogonal functions).  We refer to them as `eigenglaciers' in homage to the equivalently defined `eigenfaces' often employed in facial recognition problems \citep{sirovich1987low}.  The diagonal entries of $\Lambda$ represent the variance in the data (once again, here these are a large set of model results) explained by each of these eigenglaciers in descending order.  As such, we can simplify the representation of the data by assessing the fraction of the variance in the data still unexplained after representing it with $j$ components
\begin{equation}
    f(j) = 1 - \frac{\sum_{i=1}^j \Lambda_{ii}}{\sum_{i=1}^m \Lambda_{ii}}. 
\end{equation}
We find a cutoff threshold $c$ for the number of eigenglaciers to retain by determining $c=\max_j\in\{1,\ldots,m\} : f(j)>s $.  We set $s=10^{-4}$, which is to say that we retain a sufficient number of basis functions such that we can represent 99.99\% of the velocity variability in the model ensemble.  For the experiments considered here, $c\approx50$.  

Any velocity field that can be produced by the high-fidelity model can be approximately represented as 
\begin{equation}
    \mathcal{F}(\mathbf{m}) \approx \sum_{j=1}^c \lambda_{j}(\mathbf{m}) V_j,
\end{equation}
where $V_j$ is the $j-$th eigenglacier, and $\lambda_{j}$ is its coefficient.  The (row) vector $\lambda(\mathbf{m})$ can thus be thought of as a low dimensional set of `knobs' that control the recovered model output.  

\subsubsection{Artificial Neural Network}
Unfortunately, we do not \emph{a priori} know the mapping $\lambda(\mathbf{m})$.  In the second stage of surrogate creation, we seek to train a function $\lambda(\mathbf{m};\theta)$ with trainable parameters $\theta=\{W_l,b_l,\alpha_l,\beta_l\;:\;l=1,\ldots,L\}$ such that the resulting reconstructed velocity field is as close to the high-fidelity model's output as possible, where $L$ is the number of network blocks (see below).  For this task, we use a deep but narrow residual neural network.  The architecture of this network is shown in Fig.~\ref{fig:schematic}.  As is common for artifical neural networks, we repeatedly apply a four operation block with input $h_{l-1}$ and output $h_l$.  As input to the first block we have our parameter vector, so $h_0=\mathbf{m}$.  In each block, the first operation is a simple linear transformation
\begin{equation}
    \hat{\mathbf{a}}_l = \mathbf{h}_{l-1} W_l^T + \mathbf{b}_l,
\end{equation}
where $W_l$ and $\mathbf{b}_l$ are respectively a learnable weight matrix and bias vector for block $l$.  To improve the training efficiency of the neural network, the linear transformation is followed by so-called layer normalization \citep{ba2016layer}, which z-normalizes then rescales the intermediate quantity $\hat{a}_l$
\begin{equation}
    \mathbf{a}_l = \alpha_l \frac{\hat{\mathbf{a}}_l - \mu_l}{\sigma_l} + \beta_l,
\end{equation}
where $\mu_l$ and $\sigma_l$ are the mean and standard deviation of $\hat{\mathbf{a}}_l$, and $\alpha_l$ and $\beta_l$ are learnable layerwise scaling parameters.  Next, in order for the artificial neural network to be able to represent non-linear functions, we apply an activation
\begin{equation}
    \hat{\mathbf{z}}_l = \mathrm{ReLU}(\mathbf{a}_l),
\end{equation}
where 
\begin{equation}
    \mathrm{ReLU}(x) = \mathrm{Max}(x,0)
\end{equation}
is the rectified linear unit \citep{glorot2011deep}.  Next we apply dropout \citep{srivastiva2014dropout}, which randomly zeros out elements of the activation vector during each epoch of the training phase
\begin{equation}
  \mathbf{z}_l  = \hat{\mathbf{z}}_l \odot R,
\end{equation}
where $R$ is a vector of Bernoulli distribution random variables with mean $p$.  After training is complete and we seek to evaluate the model, we set $R:=p$, which implies that the neural network produces deterministic output, with each element of the layer output weighted by the probability that it was retained during training.  Dropout has been shown to effectively reduce overfitting by preventing complex co-adaptation of weights: by never having guaranteed access to a given value during the training phase, the neural network learns to never rely on a single feature in order to make predictions.

Finally, if dimensions allow (which they do for all but the first and last block), the output of the block is produced by adding its input
\begin{equation}
\mathbf{h}_l = \mathbf{z}_l + \mathbf{h}_{l-1},
\end{equation}
a so-called residual connection \citep{he2016identity} which provides a `shortcut' for a given block to learn an identity mapping.  This mechanism has been shown to facilitate the training of deep neural networks by allowing an unobstructed flow of gradient information from the right end of the neural network (where the data misfit is defined) to any other layer in the network.  

At the last block as $l=L$, we have that $\lambda(\mathbf{m})=\mathbf{h}_L$.  In this work, $L=5$.  $\lambda(\mathbf{m})$ is then mapped to a log-velocity field via $V$, as described above.  The complete surrogate model is thus defined as
\begin{equation}
    \mathcal{G}(\mathbf{m}) = 10^{\lambda(\mathbf{m}) V^T}.
\end{equation}
\subsection{Surrogate Training}
\begin{figure}
    \centering
    \includegraphics[width=1.0\linewidth]{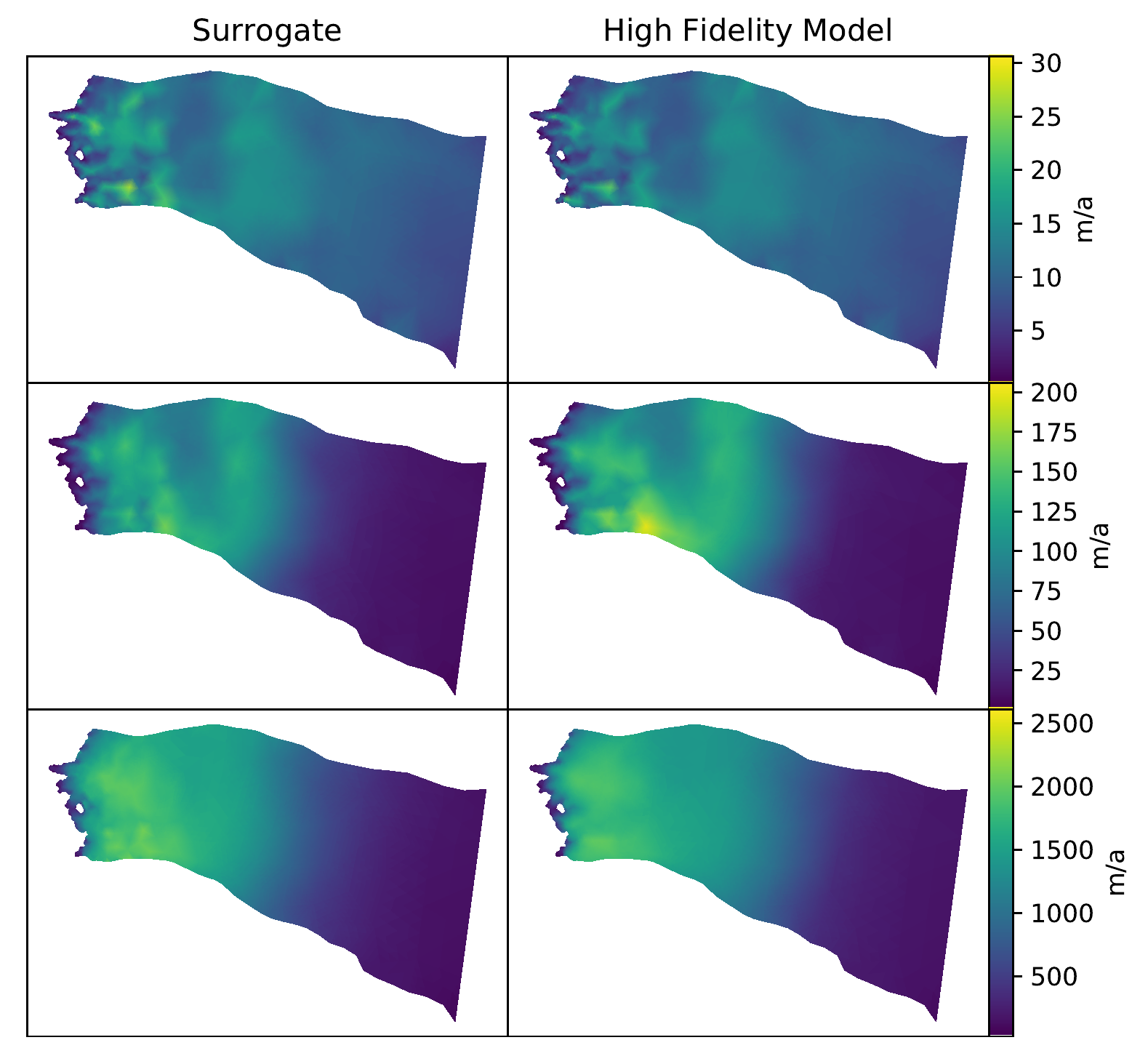}
    \caption{Comparison between emulated velocity field and modeled velocity field for three random instances of $\mathbf{m}$.  We note that these predictions are out of set: the surrogate model was not trained on these examples, and so is not simply memorizing the training data}
    \label{fig:em_vs_model}
\end{figure}
To train this model, we minimize the following objective 
\begin{equation}
    I(\theta) \propto \sum_{i=1}^m \sum_{j=1}^{n_p} \omega_{d,i} A_j \left[\log_{10} \mathcal{G}(\mathbf{m}_i;\theta)_j - \log_{10} \mathcal{F}(\mathbf{m}_i)_j\right],
    \label{eq:cost_function}
\end{equation}
where $A_j$ is the fractional area of the $j$-th grid point, and $\omega_{d,i}\in[0,1], \sum_{i=1}^m \omega_{d,i} =1$ is the weight of the $i$-th training example model error.  The former term is necessary because our computational mesh resolution is variable, and if were to simply compute the integral as a sum over grid points, we would bias the estimator towards regions with high spatial resolution.  

The model above is implemented in pytorch, which provides access to objective function gradients via automatic differentiation \citep{pazke2019pytorch}.  We minimize the objective using the ADAM optimizer \citep{kingma2014adam}, which is a variant of stochastic gradient descent.  We use a batch size of 64 input output pairs, and begin with a learning rate of $\eta=10^{-2}$, that is exponentially decayed by one order of magnitude per 1000 epochs (an epoch being one run through all of the training instances).  We run the optimization for 4000 epochs.  

The results of the surrogate training are shown in Fig.~\ref{fig:em_vs_model}.  We find that for most instances, the surrogate model produces a velocity field in excellent agreement with the one produced by the high fidelity model, with the exception of instances where the velocity fields are more than three orders of magnitude greater than observations.  Since we intend to use the surrogate for inference and such a velocity field implies that the parameters that created it are unlikely to be consistent with observations anyways, this extreme-value misfit will not influence the inference over glacier model parameters.

\subsection{Bayesian Bootstrap Aggregation}
Neural networks are known to be high-variance models, in the sense that while the high-fidelity model may exhibit a monotonic relationship between input parameters and output velocities, the neural network may exhibit high frequency `noise,' similar to that exhibited to fitting high-order polynomials to noisy data.  This noise is problematic in that it tends to yield local minima that prohibit optimization and sampling procedures from full exploration of the parameter space.  In order to reduce this variance, we introduce Bayesian bootstrap aggregation \citep{clyde2001bagging,breiman1996bagging}  (so-called bagging), in which we train the surrogate described $B$ times, with the sample weights used in Eq.~\ref{eq:cost_function} each time randomly drawn from the distribution
\begin{equation}
    \omega_{d,i} \sim \mathrm{Dirichlet}(\mathbf{1}),
\end{equation}
where $\mathbf{1}$ is a vector of ones with length $m$, the number of training instances.  

This procedure yields $B$ independent instances of $\mathcal{G}$ (with single instances hereafter referred to as $\mathcal{G}_i$), which are combined as a committee.  One way to think about this process is that the high-fidelity model is the mean of a distribution, and each ensemble member is a `data point' (a random function) drawn from that distribution.  The optimal estimate of the true mean (once again, the high fidelity model) is the sample mean of the bootstrap samples 
\begin{equation}
    \bar{\mathcal{G}}(\mathbf{m}) = \sum_{i=1}^B \omega_{e,i} \mathcal{G}_i(\mathbf{m}),
    \label{eq:surrogate_ensemble}
\end{equation}
with the weights $\omega_{e,i}\in[0,1],\sum_{i=1}^B\omega_{e,i}=1$. While this aggregation reduces predictive error (i.e. yields a better approximation to the high-fidelity model) relative to using a single model, uncertainty remains because we are approximating the true mean with the mean based on a finite number of samples.  To account for this residual uncertainty in the surrogate model, we can once again employ Bayesian bootstrapping \citep{rubin1981bayesian}.  In principle, we assume that the sample (the computed members of the bagging committee) provide a reasonable approximation to the population (all possible members of the bagging committee) when estimating variability in the mean.  In practice, this means that we model $G(\mathbf{m})$ as a random function given by Eq.~\ref{eq:surrogate_ensemble} augmented with Dirichlet distributed weights 
\begin{equation}
    \omega_{e,i} \sim \mathrm{Dirichlet}(\mathbf{1}).
\end{equation}

\section{Bayesian Inference}
We would like to infer the posterior distribution of model parameters $\mathbf{m}$ given observations $\mathbf{d}$, with the added complexity that the random surrogate described above is only an approximation to the high-fidelity model.  This can be accomplished by marginalizing over the surrogate distribution, or equivalently the bootstrap weights $\omega_e$.   
\begin{equation}
    P(\mathbf{m}|\mathbf{d}) = \int P(\mathbf{m},\omega_e | \mathbf{d}) \mathrm{d} \omega_e 
\end{equation}
Applying Bayes theorem to the right hand side, we have that
\begin{align}
                           P(\mathbf{m}|\mathbf{d})  &\propto \int P(\mathbf{d}|\mathbf{m},\omega) P(\mathbf{m},\omega_e) \mathrm{d} \omega_e \nonumber \\ 
                             &\propto \int P(\mathbf{d}|\mathbf{m},\omega) P(\mathbf{m}) P(\omega_e) \mathrm{d} \omega_e,
\end{align}
where we have used the fact that the bootstrap weights and model parameters are independent.  On the left hand side is the quantity of interest, the posterior distribution of model parameters given observations, while inside the integral, $P(\mathbf{d}|\mathbf{m})$ is the likelihood of observing the data given a set of model parameters, and $P(\mathbf{m})$ is the prior distribution over model parameters.      
\subsection{Likelihood Model}
Observations of surface velocity are reported as a field, as are the model predictions, and thus we have an infinite dimensional Bayesian inference problem \citep{bui2013computational,petra2014computational} because there are an infinite number of real-valued coordinates at which to evaluate misfit.  However, in contrast to previous works, rather than finite observations with an infinite parameter space, we have the converse, with continuous (infinite) observations and finite dimensional parameters.  To circumvent this difficulty, we propose a relatively simple approximation that can account for observational correlation and a variable grid size.  We first assume a log-likelihood of the form
\begin{equation}
    \log P(\mathbf{d}|\mathbf{m},\omega_e) \propto -\frac{1}{2} \int_{\bar{\Omega}} \int_{\bar{\Omega}'} \frac{r(x) r(x')}{\sigma(x,x')} \rho_d^2 \mathrm{d}\bar{\Omega}' \mathrm{d}\bar{\Omega},
    \label{eq:log-like}
\end{equation}
where $\rho_d$ is the data density (number of observations per square meter), $\sigma(x,x')$ is a covariance function 
\begin{equation}
    \sigma(x,x') = \sigma^2_{obs} + \sigma^2_{cor}\left(1 + \frac{d(x,x')}{2 l^2}\right)^{-1}
\end{equation}
that superimposes white noise with variance $\sigma^2_{obs}$ and rational exponential noise with variance $\sigma^2_{cor}$ and characteristic length scale $l$.  $r(x)$ is a residual function given by
\begin{equation}
    r(x) = \bar{\mathcal{G}}(x;\mathbf{m},\omega_e) - \|\mathbf{u}_{obs}\|_2(x),
\end{equation}
where $\mathbf{u}_{obs}$ is the satellite derived, annually averaged velocity field described in the Study Area section, and in which we omit writing the dependence on $\mathbf{m}$ for brevity.

Because solutions are defined over a finite element mesh, we split the integrals in Eq.~\ref{eq:log-like} into a sum over dual mesh elements $T$ in collection $\mathcal{T}$
\begin{equation}
    \log P(\mathbf{d}|\mathbf{m},\omega_e) \propto -\frac{1}{2} \sum_{T\in\mathcal{T}} \sum_{T'\in\mathcal{T}} \int_T \int_{T'} \frac{r(x) r(x')}{\sigma(x,x')} \rho^2 \mathrm{d}T' \mathrm{d}T.
\end{equation}
Finally, we make the approximation
\begin{equation}
    \int_T \int_{T'} \frac{r(x) r(x')}{\sigma(x,x')} \rho^2 \mathrm{d}T' \mathrm{d}T \approx \frac{r(x_T) r(x_{T'})}{\sigma(x_T,x_{T'})} \rho^2 A_{T'} A_T, 
\end{equation}
where $x_T$ are the coordinates of the barycenter of $T$ (the finite element mesh nodes) and $A_T$ its area.  Defining 
\begin{equation}
\mathbf{r}^T = [r(x_1),r(x_2),\ldots,r(x_N)]
\end{equation}
and 
\begin{equation}
    \Sigma^{-1} = K \hat{\Sigma}^{-1} K,
\end{equation}
where $\hat{\Sigma}_{ij} = \sigma(x_i,x_j)$ and $K=\mathrm{Diag}([\rho A_1, \rho A_2, \ldots, \rho A_N])$ yields the finite-dimensional multivariate-normal likelihood
\begin{equation}
P(\mathbf{d}|\mathbf{m}) \propto \mathrm{exp} \left[ -\frac{1}{2} \mathbf{r}^T \Sigma^{-1} \mathbf{r} \right].
\end{equation}
\subsection{Prior Distribution}
In principle, we have very little knowledge about the actual values of the parameters that we hope to infer and thus would like to impose a relatively vague prior during the inference process.  However, because the surrogate is ignorant of the model physics, we must avoid allowing it to extrapolate beyond the support of the ensemble.  One choice that fulfills both of these objectives is to use as a prior the same log-uniform distribution that we used to generate the surrogate.  However, the ensemble distribution was designed to cover as broad a support as possible without biasing the surrogate towards fitting parameter values near some kind of mode and does not represent true prior beliefs about the parameter values.  Instead, we adopt for the parameters a scaled log-Beta prior
\begin{equation}
    \frac{\log_{10} \mathbf{m} - \mathrm{Bound}_L}{\mathrm{Bound}_U - \mathrm{Bound}_L}  \sim \mathrm{Beta}(\alpha=2,\beta=2)
\end{equation}
This prior reflects our belief that good parameters values are more likely located in the middle of the ensemble, while also ensuring that regions of parameter space outside the support of the ensemble have zero probability.  

\subsection{Marginalization over $\omega_e$}
In order to perform the marginalization over bootstrap weights, we make the Monte Carlo approximation
\begin{align}
    \int P(\mathbf{d}|\mathbf{m},\omega_e) P(\mathbf{m}) P(\omega_e) \mathrm{d} \omega_e  & \approx \sum_{i=1}^N P(\mathbf{d}|\mathbf{m},\omega_{e,k}) P(\mathbf{m}),
    \label{eq:posterior}
\end{align}
with $\omega_{e,i}$ drawn as in Eq.~53, where $N$ is a number of Monte Carlo samples.  The terms in the sum are independent, and may be computed in parallel.  However they are also analytically intractable.  Thus, we draw samples from each of the summand distributions (the posterior distribution conditioned on an instance of $\omega_e$) using the MCMC procedure described below, then concatentate the sample to form the posterior distribution approximately marginalized over $\omega_e$.  The marginalization of the posterior distribution in this way is similar to BayesBag \citep{buhlmann2014discussion,huggins2019using}, but with bootstrap sampling applied over models rather than over observations.      

\subsection{Manifold Metropolis Adjusted Langevin Algorithm}
As is typical for Bayesian inference, the posterior distributions $P(\mathbf{m}|\mathbf{d},\omega_e)$ are intractable, and we turn to Markov Chain Monte Carlo (MCMC) methods to draw samples \citep{kass1998markov}.  MCMC methods operate by performing a random walk in parameter space, with candidate for the next position $\hat{\mathbf{m}}_{t+1}$ determined according to a proposal distribution $Q(\cdot|\cdot)$ \begin{equation}
\hat{\mathbf{m}}_{t+1} \sim Q(\hat{\mathbf{m}}_{t+1}|\mathbf{m}_t).
\end{equation}
A given candidate parameter vector is accepted or rejected according to its posterior probability relative to the current position in parameter space:
\begin{equation}
a = \min\left(1,\frac{P(\hat{\mathbf{m}}_{t+1}|\mathbf{d}) Q(\mathbf{m}_t|\hat{\mathbf{m}}_{t+1}}{P(\mathbf{m}_t|\mathbf{d})Q(\hat{\mathbf{m}}_{t+1}|\mathbf{m}_t)}\right),
\end{equation}
where $a$ is the probability of acceptance.  If a proposal is accepted, then $\mathbf{m}_{t+1} := \hat{\mathbf{m}}_{t+1}$; otherwise, $\mathbf{m}_{t+1}:=\mathbf{m}_t$.  In the limit as $t\rightarrow \infty$ (and under some restrictions on the proposal distribution), the set of samples produced by this procedure converges to the true posterior distribution $P(\mathbf{m}|\mathbf{d})$.

Because of the potential for highly correlated parameters, a simple application of (for example) the Metropolis-Hastings algorithm (which utilizes an isotropic Gaussian distribution centered around the current position as a proposal distribution) is unlikely to efficiently explore the space.  However, because of the availability of automatic differentiation for the surrogate model we have easy access to the gradient of the log-posterior.  This allows for a sampler that can efficiently steer itself towards probable regions of parameter space.  Furthermore, because this inference problem is low dimensional, it is straightforward to compute the gradient of the gradient (i.e. the Hessian matrix), which allows for an efficient scaling of the proposal distribution.  

One method which allows us to capitalize on this availability of derivatives is the manifold-Metropolis Adjusted Langevin Algorithm \citep[mMALA,][]{girolami2011riemann}.  mMALA operates as described above, but with proposal distribution given by 
\begin{equation}
    Q(\hat{\mathbf{m}}_{t+1}|\mathbf{m}_t) = \mathcal{N}(\mathbf{m}_t - \Delta \hat{H}^{-1} \nabla \log P(\mathbf{d}|\mathbf{m}_t,\omega_e), 2 \Delta \hat{H}^{-1}),
\end{equation}
where $\hat{H}$ is a an approximation to the Hessian that is regularized to be positive definite.  This method is very similar to the stochastic Newton MCMC method proposed by \citet{petra2014computational}, but with the use of an analytical (rather than numerically approximated) Hessian and a generalization to step size $\Delta\neq 1$, which we have found to be critical for numerical stability.   

\begin{figure}
    \centering
    \includegraphics[width=0.70\linewidth]{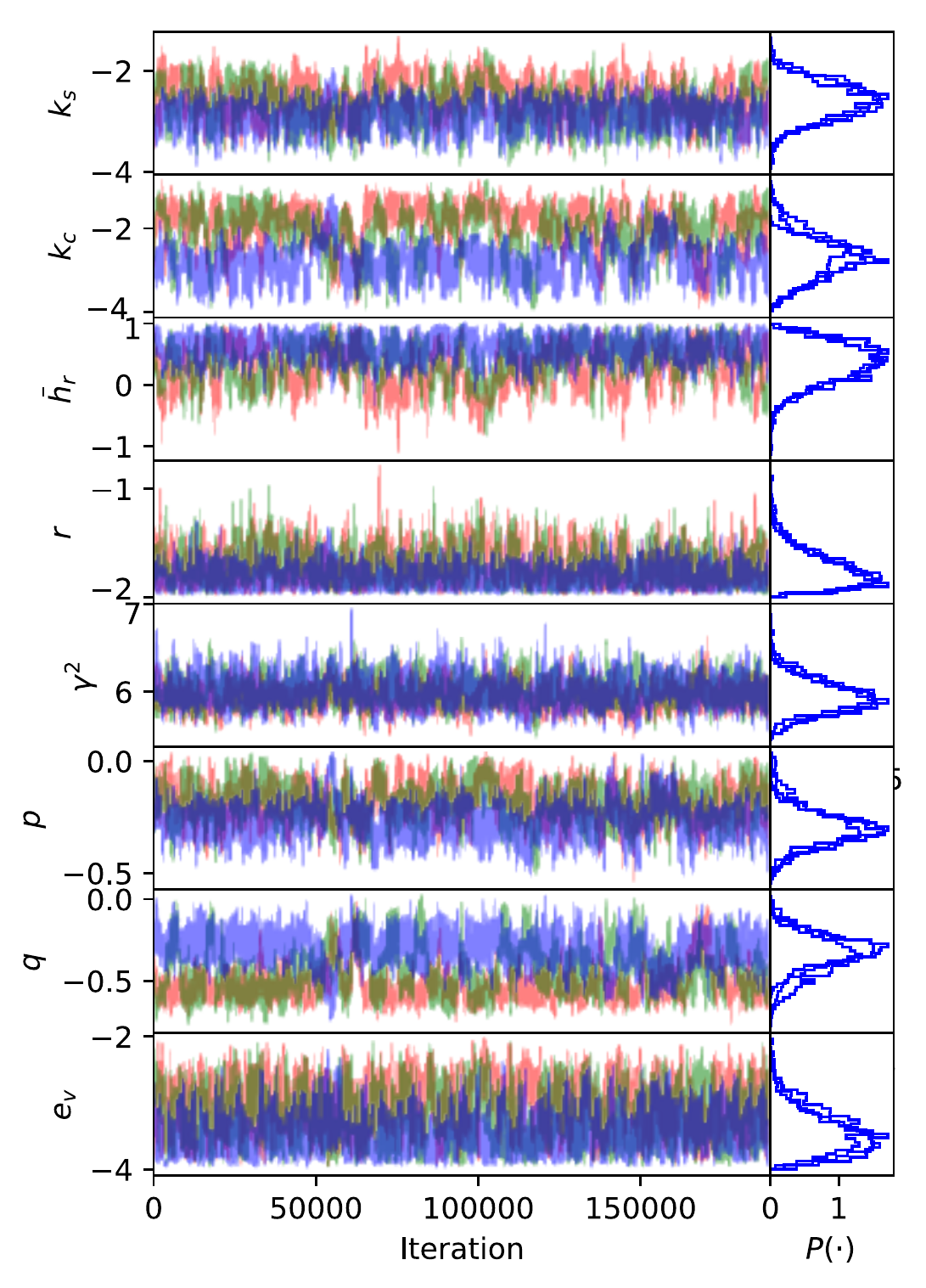}
    \caption{Three Markov chains over the base-10 logarithm of parameter values (Left, RGB), each for a different random value of $\omega_e$.  The ``fuzzy caterpillar'' pattern indicates good mixing.  The right plot shows histogram of the blue sample, after being divided into three disjoint sub-chains.  The very similar histograms indicate a converged chain. }
    \label{fig:chains}
\end{figure}

For each summand in Eq.~\ref{eq:posterior}, we initialize the sampler from the maximum a posteriori point, which is computed via Newton's method (again, trivial to implement due to the availability of the Hessian), initialized from a random draw from the prior distribution.  We run the sampler for $2\times10^5$ iterations, with a step size selected by a simple moving average scheme that aims to keep the sampler's acceptance rate at approximately 0.56, the theoretically optimal value for mMALA \citep{roberts2001optimal}.  Performing this process for each summand leads to $N=100$ randomly initialized chains, which helps to minimize the likelihood that any individual chain is stuck in a local minimum.  We discard the first $10^4$ samples as burn-in.  The resulting chains are shown parameter-wise in Fig.~\ref{fig:chains}.  From a qualitative perspective, the chains exhibit good mixing, as indicated by the ``fuzzy caterpillar'' pattern.  We ensure that the distributions are approximately stationary by dividing each chain into thirds, and overlaying the resulting histograms; we find that the histograms are very similar, indicating approximate MCMC convergence.  Remaining MCMC error is further ameliorated by taking the expectation over independent chains.  
\subsection{Posterior Distribution}
\begin{figure*}
    \centering
    \includegraphics[width=\linewidth]{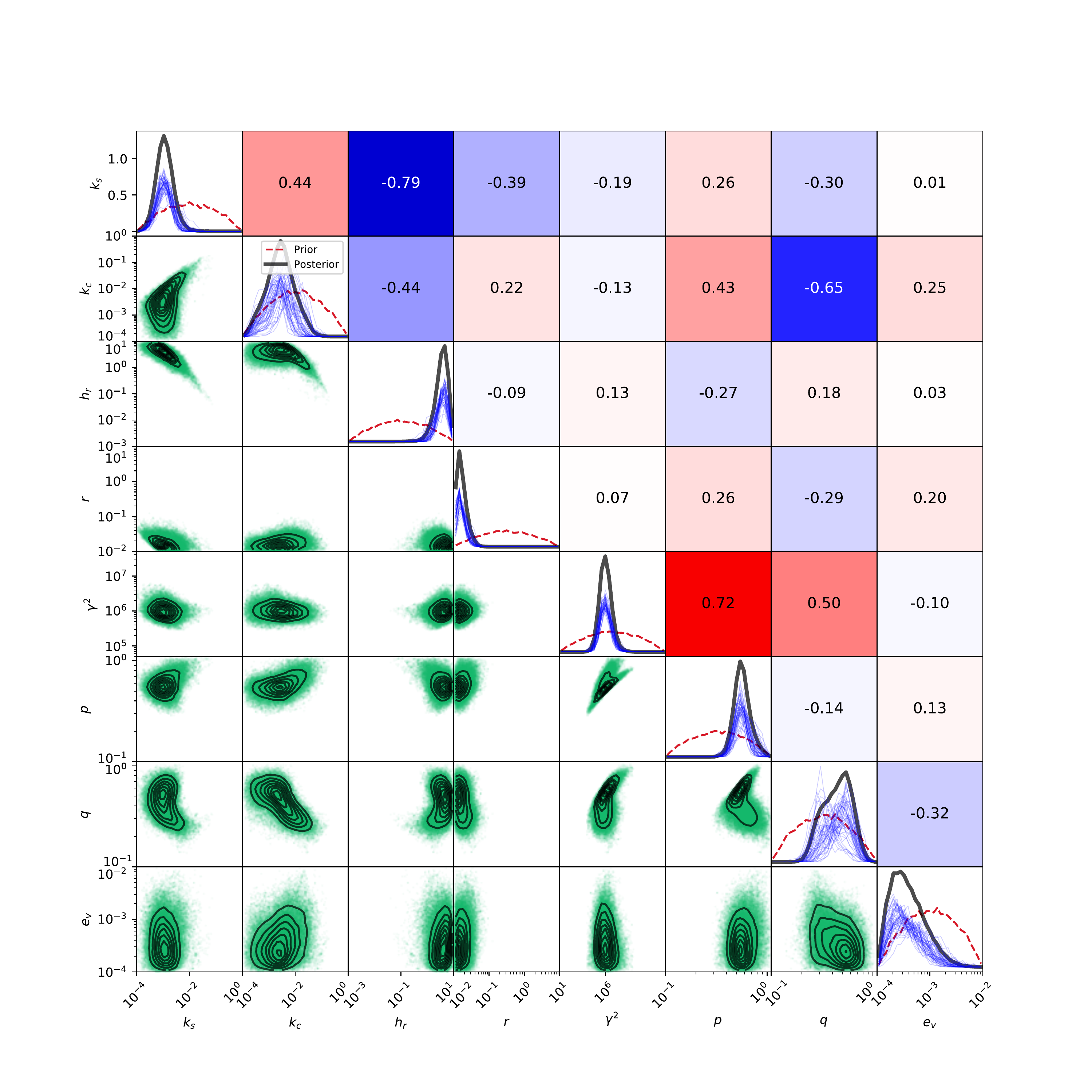}
    \caption{Posterior distributions.  (Diagonal) marginal distributions for the posterior (black) and prior distribution (red), with BayesBag posteriors in blue (at half scale for clarity).   (Below diagonal) pairwise marginal distributions illustrate correlation structure between parameters.  (Above diagonal) correlation coefficient for each pair of parameters, with red and blue corresponding to positive and negative correlations, respectively.}
    \label{fig:scatter}
\end{figure*}

The diagonal entries in Fig.~\ref{fig:scatter} show the prior and posterior marginal distributions for each of the eight parameters in $\mathbf{d}$.  One immediate observation is that the posterior distributions for all parameters exhibit a significantly reduced variance relative to the prior distribution.  This implies that surface velocity information alone conveys information not only about the sliding law, but also about the parameters of the hydrologic model.  
\subsubsection{Hydrology Parameters}
We find that the hydraulic conductivity has a mean value of approximately $k_s=10^{-3}$, but with a 95\% credibility interval of around an order of magnitude in either direction.  Unsurprisingly, this parameter exhibits a strong negative correlation with characteristic bedrock bump height $h_r$: because flux through the inefficient system is a function that increases with both transmissivity and cavity height, an increase in one term can be compensated for by the other.  Interestingly, bedrock bump heights most consistent with observations are on the order of meters.  We emphasize that this does not imply that average cavity heights are on the order of meters; in fact, the model typically predicts average cavity thickness on the order of 10s of centimeters (See Fig.~\ref{fig:predictive_hydrology}).  Rather, this result implies that the model should never reach $h=h_r$, at which point the opening rate begins to decouple from velocity.  Nonetheless, this rather large bedrock asperity size introduces the \emph{potential} for very large cavities to form.  This tendency is offset by a very low bump aspect ratio $r$, which tends to be less than $0.1$.  Conditioned on the hypothesized physics, the observations indicate an inefficient drainage system formed around large and low-slope bedrock features. 

A particularly interesting feature of these results is found in the distribution over channel transmissivity $k_c$.  Of the various parameters governing subglacial hydrology, this one is the most poorly constrained.  As shown in Fig.~\ref{fig:predictive_hydrology}, there are a number of drainage configurations that are consistent with observations, from essentially negligible to extensive.  This insensitivity means that a broad array of channel conductivities are possible, and also implies that more work is needed either to quantify the influence of the efficient system on ice dynamics or to directly observe the channel network in order to constrain this value for prognostic modelling.  

The englacial porosity $e_v$ controls the speed at which the hydrologic head changes in response to alterations in flux or forcing.  We find that this parameter is relatively poorly constrained by observations relative to prior assumptions.  This is not surprising: we would expect the influence of this parameter to primarily manifest itself by controlling the rate of change of water pressure and hence velocity.  Since we only consider time-integrated quantities here, this characteristic is not well constrained.  Nonetheless, this work suggests a porosity that is on the lower end of the plausible spectrum of values.  This indeterminacy also motivates the potential utility for time dependent inversion (see Discussion).   

\subsubsection{Sliding Law Parameters}
$\gamma^2$ exhibits a strong positive correlation with $p$.  This is simply the result of an increase in $p$ yielding an immediate decrease in the sliding law pressure term (which is typically less than unity), and thus a commensurate increase in $\gamma^2$ will yield a similar sliding velocity.  This is also true (though to a much lesser extent) of $\gamma^2$ and $q$.  $\gamma^2$ is strongly constrained by observations, as it sets the scale of glacier velocity, which is directly observable.  

The pressure exponent $p$ has a median value of approximately $p=0.5$, with a relatively small variance.  Similarly, the sliding law exponent $q$ also has a median value of approximately $q=0.5$, but with a significantly larger spread.  This spread is distinctly non-Gaussian.  Indeed, based on the curvature evident in the joint distributions between $q$ and most other variables, it seems that the distribution over $q$ is the superposition of two overlapping distributions, one associated with a value of $q$ closer to 0.6 (which agrees well with \citet{Aschwanden2016}, and the other (somewhat less probable) mode around $q=0.2$.  This latter secondary mode implies that pseudoplasticity may also be an appropriate bed model.  It seems possible that this `indecision' on the part of the sampler implies that different regions of the glacier might be better fit by different sliding laws, an unsurprising result if some regions are underlain by till and some directly by bedrock.  These two modes also lead to different preferred hydrologic parameters: in the pseudo-plastic mode, we see somewhat greater transmissivities, and a somewhat smaller characteristic asperity size.  

\subsection{Posterior Predictive Distribution}
\begin{figure}
    \centering
    \includegraphics[width=0.6\linewidth]{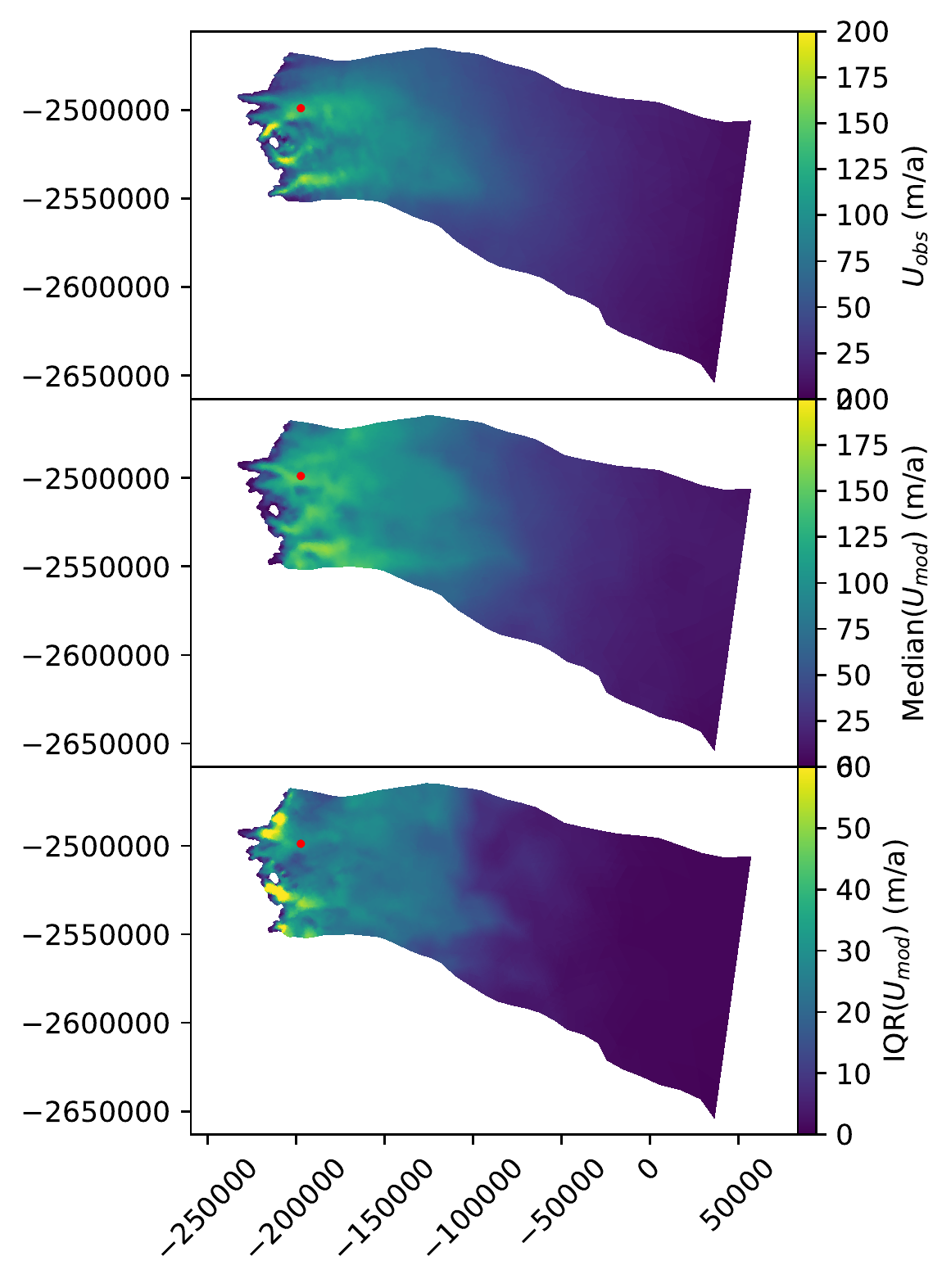}
    \caption{Posterior predictive distribution.  (Top) Observed velocity for study site.  (Middle) Median of predicted velocity fields computed by running the high-fidelity model on samples from the posterior distribution from Fig.~\ref{fig:scatter}.  (Bottom) Interquartile range of velocity posterior predictive distribution.  The red dot is the location at which a time series is extracted for Fig.~\ref{fig:time_series}.}
    \label{fig:predictive}
\end{figure}
The inference above was performed using a surrogate model, and while the surrogate reproduces predictions from the high fidelity model in the large majority of circumstances, we have yet to ensure that samples from the posterior distributions inferred using the surrogate produce velocity fields that are consistent with observations when fed back into the high-fidelity model.  We note that we do not expect perfect correspondence to observations: the model is necessarily a substantial simplification of a highly complex and heterogeneous physical system.  Rather, we seek to verify that a) the surrogate does a good job of reproducing model predictions in regions of high posterior probability, and b) that samples drawn from both the posterior distribution lead to velocity predictions that are consistent with observations to the extent that this is possible.

We selected 256 random samples from the posterior distribution shown in Fig.~\ref{fig:scatter}, and ran the high fidelity model with these parameter values.  Fig.~\ref{fig:predictive} shows the mean velocity field as well as the interquartile range, along with the observed velocity.  We find that the model fits the observations reasonably well, with an appropriate pattern of fast flow in the outlet glaciers and slow flow in the interior.  The transition between these two regimes near the equilibrium line altitude (ELA) is also well-captured by the model.  However, the model produces velocity predictions that are somewhat more diffuse than observations, and also fails to match the high-velocities evident in some steep marginal areas.   The spread in model predictions is consistent with the imposed observational uncertainty, with an IQR of between 20 and 30 over most of the ice sheet below the ELA.  Above the ELA, the predicted spread is \emph{lower} than the observational uncertainty in slow flowing regions, indicating that the model is less sensitive to parameter choice in this region than the faster flowing areas downstream.  Nonetheless, sliding still makes up approximately 80\% of the modelled (and presumably observed) surface velocity there.  Conversely, the model error induced by the surrogate leads to somewhat higher spread in some fast flowing regions near the margin, likely due to these being the places where significant non-linearity in the model (e.g. channelizations, reaching the ``elbow'' of the sliding law, etc.) occur, and hence are more challenging to emulate.  

\begin{figure}
    \centering
    \includegraphics[width=0.6\linewidth]{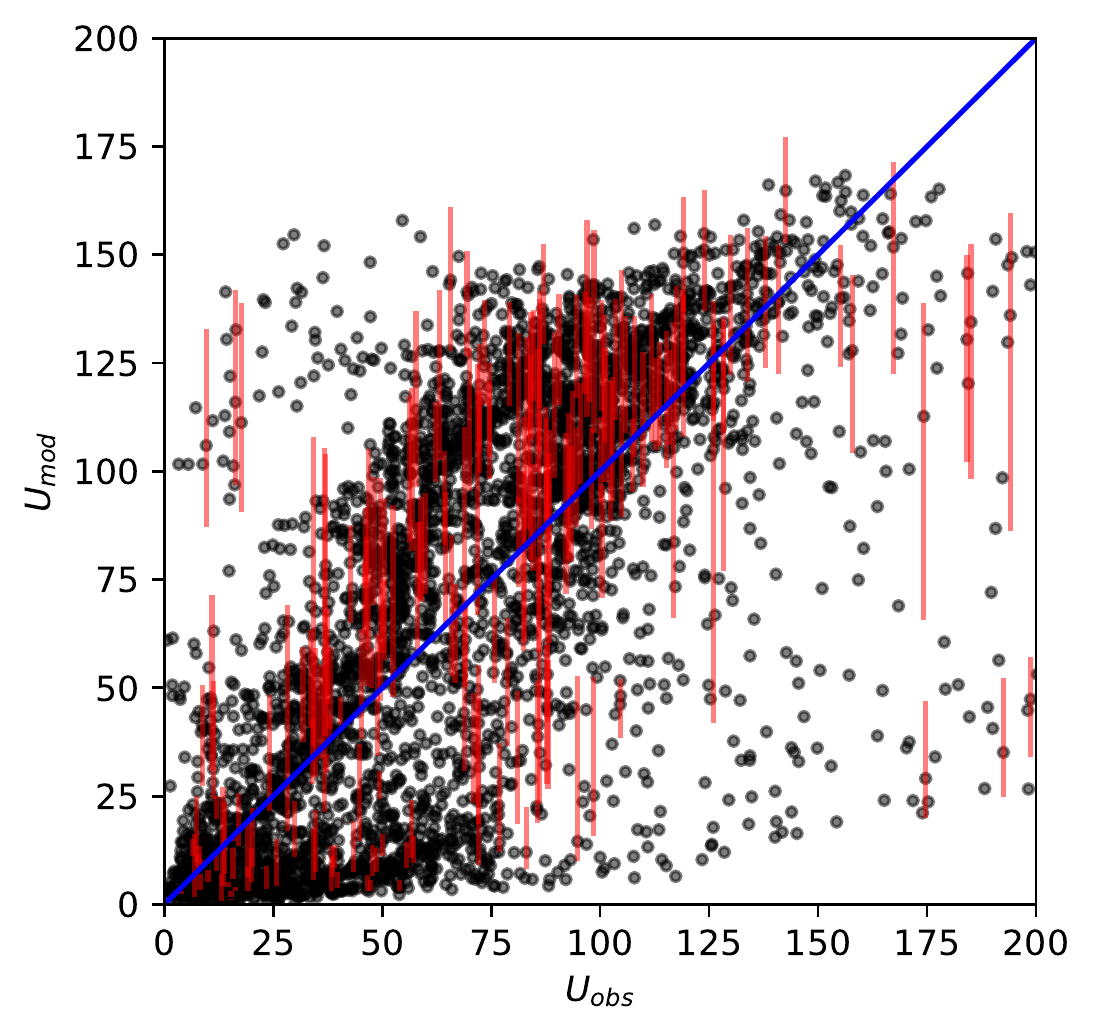}
    \caption{Observed versus ensemble modelled velocities.  Modelled spread is given by red lines, plotted for every twenty points.  Blue line gives a one-to-one correspondence.  Median Bayesian $R^2=0.6$.}
    \label{fig:gof_plot}
\end{figure}

It is also useful to establish the degree to which the optimized model explains the observation.  Fig.~\ref{fig:gof_plot} shows the velocity observations versus predictions in the form of a scatter plot, as well as the model's predictive spread.  Clearly, the model carries substantial predictive power, however there is also substantial variability around the 1:1 line.  One simple goodness-of-fit metric is the Bayesian $R^2$ \citep{gelman2019r}, which measures the variance in model predictions relative to the variance of model predictions plus the variance of the residuals.  For a model that perfectly models the data, $R^2=1$, and for values less than unity $R^2$ quantifies the fraction of data variance explained by the model.  Here, we find a median value of $R^2=0.6$, indicating that the model explains 60\% of the variance in the observations.  Taking this number and the results in Fig.~\ref{fig:gof_plot} together, particularly given the non-Gaussianity of the residuals, we think that the model presented here is underparameterized: a model that allows for some spatial variability in basal conditions would likely fit the data better, and would also be conceptually justifiable, given that different regions of the bed have different geology and sediment cover.  However, determining how to parameterize this variability without a wholesale return to the difficulties associated with spatially explicit traction coefficients remains a challenge.

\subsubsection{Hydrologic configuration}
\begin{figure*}
    \centering
    \includegraphics[width=\linewidth]{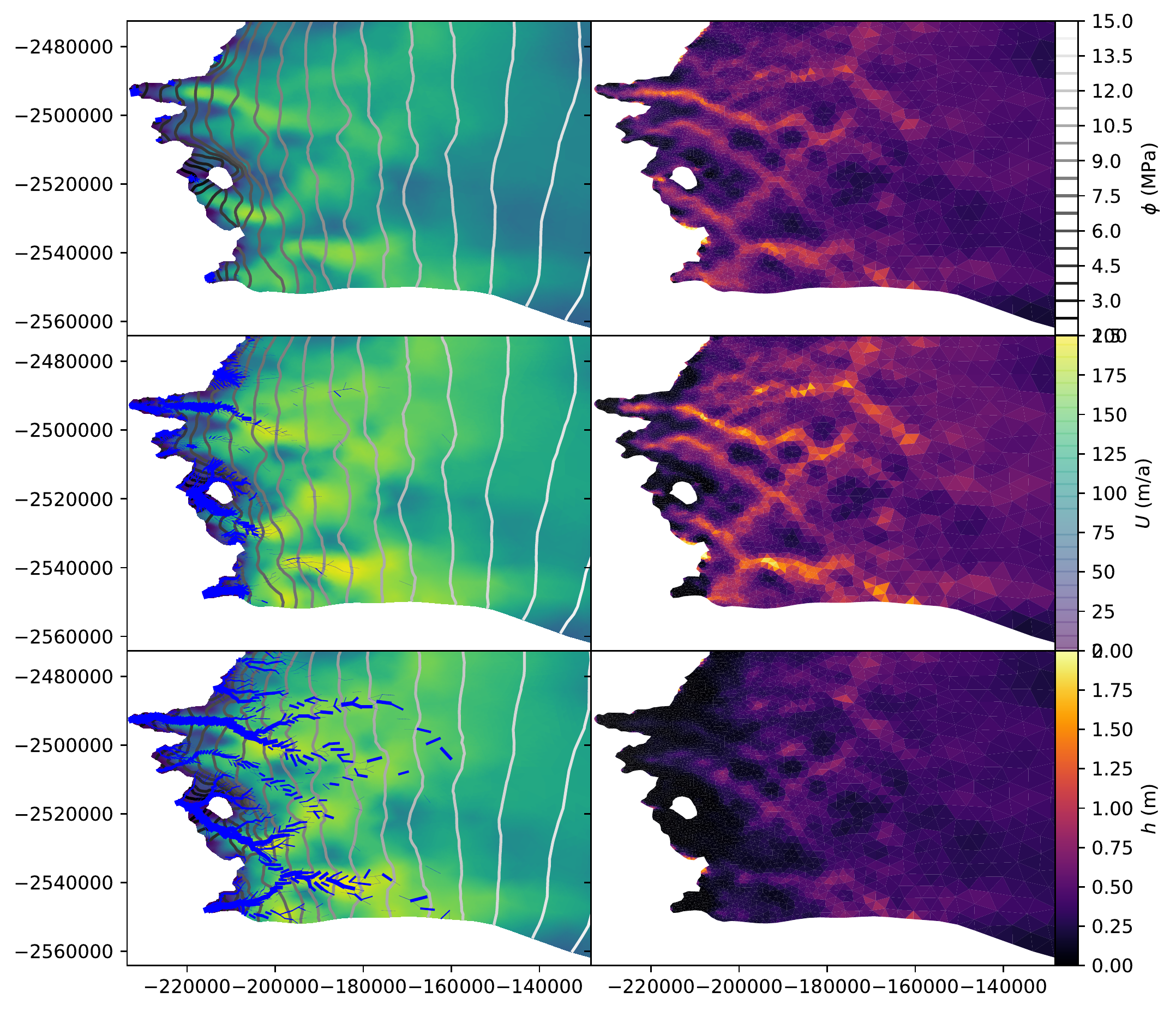}
    \caption{(Left) Annual average configuration of channels for the simulation according to the 16th (top), 50th (middle), and 84th (bottom) quantile of annually integrated channelized system flux.  The widest blue line is approximately 300m$^3$s$^{-1}$ while the smallest visible lines are 10$^{-2}$m$^3$s$^{-1}$.  Contours show the hydropotential.  (Right) associated distributed water layer thickness fields.}
    \label{fig:predictive_hydrology}
\end{figure*}
While our surrogate model does not provide direct access to the state variables of the hydrologic model, the posterior predictive samples do.  In Fig~\ref{fig:predictive_hydrology}, we show the hydraulic potential, channel flux, and subglacial cavity size for a weakly, moderately, and strongly channelized posterior sample, all of which produce velocities that are (more or less) equivalently consistent with observations.  In the weakly channelized case, large channels occur only near the terminus, where large upstream areas and low overburden pressures allow very large but highly localized channels to form.  We note that this low channelization case produces a spacious distributed system, with $h$ frequently reaching 1m in areas of convergent topography (e.g. the bottom of troughs).  A much more well-developed channelized system develops in the moderately channelized sample.  However, the inefficient drainage system magnitude remains similar, indicating that despite its greater extent, the channelized system transports relatively little water.  Conversely, in the most channelized model run, channels extends nearly all the way to the ELA.  The resulting distributed system configuration has much less capacity, with the average cavity size rarely exceeding 0.25 m.

\subsubsection{Temporal Changes in velocity}
\begin{figure}
    \centering
    \includegraphics[width=0.6\linewidth]{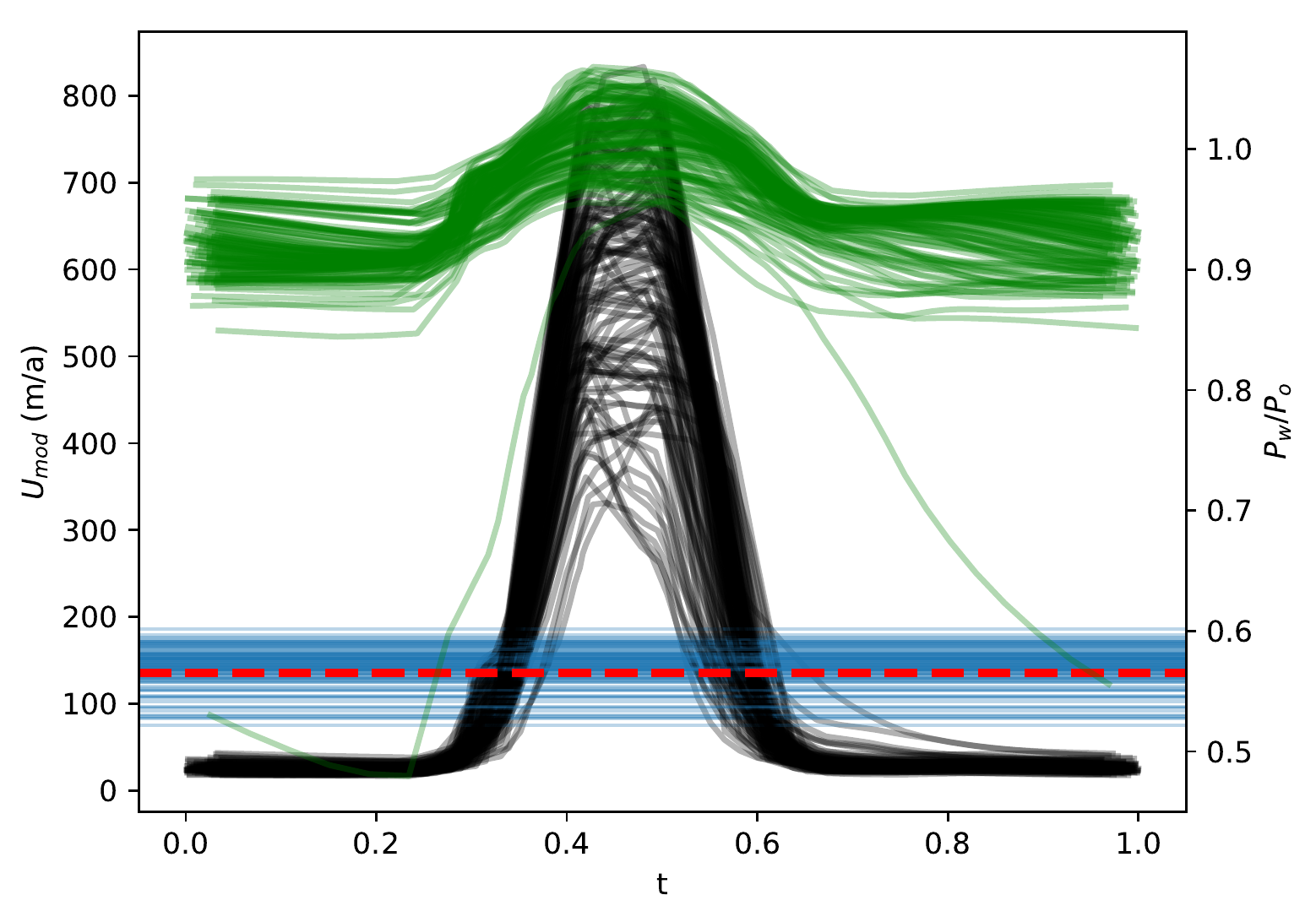}
    \caption{Time series of velocity (black) over a single year at the red point in Fig.~\ref{fig:predictive}, modelled annual averages (blue), observed annual average (red), and fraction of overburden (green).}
    \label{fig:time_series}
\end{figure}

While we constructed the surrogate model and inferred parameters based on time-averaged velocities, the underlying model is still time-dependent and it is of substantial interest to examine the time-dependent behavior of the model.  Fig.~\ref{fig:time_series} shows the ice sheet's speed and water pressure as a fraction of overburden in the middle of Isunnguata Sermia, coincident with the red dot in Fig.~\ref{fig:predictive}.  While we find similar qualitative behavior in each simulation, namely an increase in water pressure associated with the onset of meltwater in the spring and a coincident increase in velocity, the peak velocity and speedup duration varies significantly between simulations.  This spread in behavior occurs despite annual average velocities that are consistent with observations conditioned on the uncertainty assumptions stated above.  This spread is most acutely driven by uncertainty in the englacial porosity $e_v$, which plausibly varies by nearly two orders of magnitude, and controls the water pressure rate of change.

In nearly all simulations, water pressure is uniformly high throughout the year, reaching or exceeding overburden pressure during the meltwater season.  This uniformly high pressure is consistent with observations for this reason.  However, the annual pattern of velocity remains \emph{inconsistent} with the observational record \citep[e.g][]{andrews2014direct,moon2014distinct}, in particular the lack of a significant winter speed-up.  One important future line of inquiry that we are currently undertaking is whether the current model (or any currently proposed hydrologic model) can replicate this time-varying field for \emph{any parameter combination}.  If so, then the posterior parameter variance will likely be reduced substantially.  However, to answer this question in the negative would call into considerable question the utility of hydrologic models for glaciological modelling.

\section{Discussion}

\subsubsection{Model Selection}
To paraphrase \citet{box1987empirical}: ``All models are wrong, but some are useful.''  Despite the relative robustness of the Bayesian framework here, its ability to quantify parametric uncertainty, and the model's encouraging ability to reproduce many salient features of the velocity observations, we remain skeptical of drawing conclusions that are too certain.  This skepticism emerges primarily from the issue of model misspecification: it is almost certainly the case that neither the hydrologic model nor the chosen sliding law (nor even the first-order ice dynamics) are a wholly appropriate approximation of the true physics.  This is clearly seen in Fig.~\ref{fig:gof_plot}, which indicates that the residuals between the predicted and observed velocities possess systematic (rather than random) biases.  As such, the model is wrong, but is it useful?  We argue that this work represents a first step towards a defensible mechanism of predicting glacier sliding into the future.  However, the physics simulated here are only one possibility, and perhaps not the best possibility.  As such, one useful next step towards the goal of a prognostic sliding law would be to repeat the procedure presented here with a variety of candidate models, and to use a formal model selection criterion such as Akaike's information criterion \citep{akaike1998information} 
\begin{equation}
    AIC = 2k - 2\log P(\mathbf{d}|\mathbf{m}),
\end{equation}
which estimates the relative information loss of a set of candidate models with respect to the true data generating process, to select between them.  Indeed, we can do this very simply for the model presented here and, for example, an unregularized inversion of basal traction of the type popularized in \citet{macayeal1993tutorial}.  In the above, $k$ is the number of parameters, which in the case of this work is $k=9$ (including the data variance).  In the spatially varying inversion, $k=4042$, which is the number of grid cells plus one.  In the work presented above, the log probability at the \emph{a posteriori} most probable parameter estimate is (to a constant that cancels when comparing AIC between two models) $\log P(\mathbf{d}|\mathbf{m})\propto -74$.  In the case of the spatially varying inversion, the log likelihood is effectively zero, representing a nearly perfect fit to the data.  Thus we have $AIC\approx 166$ for the model presented here, and $AIC\approx 8042$ for a spatially varying inversion (although this number will decrease substantially in the presence of regularization, which induces a spatial covariance that decreases the number of effective parameters).  Thus, while the model presented here does not fit the data as well, this disadvantage is more than offset by its simplicity with respect to minimizing the loss of information relative to a perfect model of glacier physics.  

Nonetheless, it is unlikely that the model presented here is the optimal one.  We intent to explore this question systematically in the future by examining both alternative hydrologic and sliding parameterizations, as well as (re-)introducing spatially varying parameters in such a way that a model selection criterion such as AIC is optimized.  The framework suggested here provides a consistent methodology for coupled model optimization that can be applied to any model configuration, without the need for the implementation of challenging numerical techniques such as time dependent adjoints.

\subsubsection{Including time-dependent observations}
Another important consideration is that we use observations that are averaged over the year, thus likely discarding important information contained in time rates of change and temporal patterns.  Fortunately, the procedure presented here is easily amenable to time dependent inversion.  The only substantive difference is in the construction of the surrogate (rather than train a network to predict the coefficients of the eigenglaciers presented in Fig.~\ref{fig:basis_functions}, these basis functions must be explicit in time as well) and the likelihood function (which must now include observations at different points in time and also explicitly model spatio-temporal covariance).  

\subsubsection{Supplementary datasets}
In addition to time-varying data, it will also be important to augment velocity observations with other measurements.  In particular, including borehole measurements of water pressure would likely yield a much smaller admissible parameter space by constraining the rate of change in pointwise storage in the coupled sub-/englacial hydrologic system.  Similarly, radar derived estimates of channel extent \citep{livingstone2017paleofluvial} would provide a statistical target for determining which of the samples presented in Fig.~\ref{fig:predictive_hydrology} is most consistent with reality.  The Bayesian framework offers a natural mechanism for incorporating diverse observations into the likelihood model, and the wide availablity of such observations represents a major avenue for improvement in parameter estimation for sliding prediction.  

\subsubsection{Spatial generalization}
Finally, it remains to be seen whether the parameter distributions inferred here are transferable to other parts of Greenland.  It stands to reason that parameters that likely depend on the underlying geology, such as average asperity height $\bar{h}_r$, the ratio of asperity height to spacing $r$, and the traction coefficient $\gamma^2$ should vary across Greenland, while parameters that are more intrinsic to the ice configuration, such as hydraulic conductivities, sliding law exponents, and englacial porosity should remain close to constant.  At the very least, this work supports the notion that when parameters vary across space, it is possible that they may do so at geologically relevant spatial scales while still maintaining good fidelity to observations.

\section{Conclusions}
We developed a coupled model of subglacial hydrology and glacier flow, and used it to infer the posterior probability distribution of eight key model parameters.  Because the model is computationally expensive, this inference was non-trivial.  We first had to run a large ensemble of parallel model runs, with ensemble members constructed by sampling from the space of admissible parameter combinations.  We then used the resulting samples to train an artificial neural network to act as a surrogate for expensive model physics.  Because the neural network was not a perfect reproduction of model physics, we introduced a double bootstrap aggregation approach to both smooth the surrogate's response to different parameters, and also to robustly account for model error.  With the surrogate in hand, we ran a Markov Chain Monte Carlo method to draw samples from the posterior distribution given an observed annual average velocity field.  We found that the velocity observation provided substantial information about all of the model parameters relative to a prior distribution, though some were more strongly constrained than others.  In particular, we found that both transmissivity of the subglacial conduit network and the englacial porosity remain highly uncertain, and this uncertainty leads to a qualitative variety of solutions that are consistent with observations.  Nonetheless, we find that this eight parameter model can account for 60\% of variance in the observational dataset, and produces velocity fields that are spatially consistent with observations.   

\section{Acknowledgements} 
We acknowledge Ruth Mottram for providing the HIRHAM surface mass balance fields.  We thank Mauro Werder who provided key insights when reimplementing GlaDS in FEniCS.  A.A., M.A.F., and D.J.B. were supported by NASA Cryosphere Grant NNX17AG65G.  

\bibliography{references}   
\bibliographystyle{apalike}  
\end{document}